\documentclass[traditabstract]{aa}

\usepackage{natbib}
\usepackage{color}
\usepackage{appendix}
\usepackage{rotating} 
\usepackage{lscape} 

\begin{document}

\title{The Wendelstein Calar Alto Pixellensing Project (WeCAPP): \\the M31 Nova catalogue}

\author{
C.-H.~Lee\inst{\ref{inst1}} 
\and A.~Riffeser\inst{\ref{inst1}} 
\and S.~Seitz\inst{\ref{inst1},\ref{inst2}} 
\and R.~Bender\inst{\ref{inst1},\ref{inst2}} 
\and J.~Fliri\inst{\ref{inst3}} 
\and U.~Hopp\inst{\ref{inst1},\ref{inst2}} 
\and C.~Ries\inst{\ref{inst1}} 
\and O.~B\"arnbantner\inst{\ref{inst1}} 
\and C.~G\"ossl\inst{\ref{inst1}}}

\institute{University Observatory Munich, Scheinerstrasse 1, 81679 M\"unchen, Germany\label{inst1} 
\and Max Planck Institute for Extraterrestrial Physics, Giessenbachstrasse, 85748 Garching, Germany\label{inst2}
\and GEPI, CNRS UMR 8111, Observatoire de Paris, 92195 Meudon, France\label{inst3}
}

\offprints{chlee@usm.lmu.de}
\date{Received / Accepted}

\abstract{
We present light curves from the novae detected in the long-term, M31 monitoring WeCAPP project.
The goal of WeCAPP is to constrain the compact dark matter fraction of the M31 halo with microlensing observations. 
As a by product we have detected 91 novae benefiting from the high cadence and highly sensitive 
difference imaging technique required for pixellensing. We thus can now present the largest CCD and optical 
filters based nova light curve sample up-to-date towards M31. We also obtained thorough coverage of the 
light curve before and after the eruption thanks to the long-term monitoring.
We apply the nova taxonomy proposed by Strope et al. (2010) to our nova candidates
and found 29 S-class novae, 10 C-class novae, 
2 O-class novae and 1 J-class nova. We have investigated the universal decline law advocated 
by Hachichu and Kato (2006) on the S-class novae. In addition, we correlated 
our catalogue with the literature and found 4 potential recurrent novae. Part of our catalogue has 
been used to search for optical counter-parts of the super soft X-ray sources detected in M31 (Pietsch et al. 2005). 
Optical surveys like WeCAPP, and coordinated with multi-wavelength 
observation, will continue to shed light on the underlying physical mechanism of novae in the future.
}


\maketitle

\section{Introduction}
Classical novae span a subclass of cataclysmic variables, consisting of a white dwarf which interacts 
with a late-type companion star. The companion loses its mass through Roche lobe overflow, 
forming an accretion disk around the white dwarf. The mass transfer from the companion induces 
thermo-nuclear runaway (TNR) onto the surface of the white dwarf, which leads to the nova eruption. 

Novae are important in several aspects. First of all, they have the potential to serve as standard candles 
of extra-galactic distance indication. This is due to the relation between the maximum luminosity 
of the light curve and the rate of decline. \cite{1929ApJ....69..103H} first noticed that brighter novae are 
prone to steeper decline. The empirical `Maximum Magnitude versus Rate of Decline' (MMRD) relation 
was further investigated by \cite{1936PASP...48..191Z} and studied quantitatively by \cite{1945PASP...57...69M} 
and \cite{1956AJ.....61...15A}. The theoretical foundation for MMRD relation is laid down
by \cite{1981ApJ...243..926S} and further revised by \cite{1992ApJ...393..516L}.

Novae could also shed light on the underlying stellar population of the environment. For example, 
\cite{1995ApJ...452..704D} proposed that fast novae ($t_2 < 12$ days) are related to stars 
belonging to Population I with relatively massive white dwarfs, while slow novae are associated to Population
II stars and have less massive white dwarfs.

In addition, novae play a role in the galactic abundances. Novae have been considered as major sources of 
galactic $^{13}$C, $^{15}$N and $^{17}$O, and minor contributors to $^7$Li, $^{9}$F and $^{26}$Al. However, 
novae hardly contribute to the overall galactic metallicity compared to supernovae or AGB stars, 
because only 10$^{-4}$ to 10$^{-5}M_{\odot}$ are ejected per nova outburst \citep{2007JPhG...34..431J}.

Recurrent novae are also regarded as possible supernovae progenitor candidates 
\citep[see e.g. ][ and reference therein.]{2010ApJS..187..275S}. 
The fundamental question is whether recurrent novae accumulate 
enough mass onto the central white dwarf envelope and turn into supernovae progenitors even after  
several novae explosions. 

Last but not the least, novae are main contributors to the class of super soft X-ray sources (SSS). 
\cite{2005A+A...442..879P} searched 
for X-ray counterparts of the optical novae in M31, and found that novae are major sources of soft 
X-ray emission. The SSS phase can provide us with information on the white dwarf mass, the ejected 
and burned mass in the outburst \citep[e.g.][]{2010AN....331..187P}. 

Due to the interstellar extinction in the Galactic disk, we can only observe 
a small fraction of the Galactic novae that erupt each year \citep{1997ApJ...487..226S}.
Thus, we need to take 
into account rather large (and likely uncertain) corrections for incompleteness when determining the 
spatial distribution or estimation of the Galactic nova rate. In such case, M31 is an ideal target for 
a novae survey because novae are still bright enough to be observed ($m_R$ $<$ 20 mag) and it is possible 
to cover the entire M31 galaxy within several pointings.  
  
Novae monitoring campaigns towards M31 can be dated back to the pioneering work done by Hubble in 
1920s \citep{1929ApJ....69..103H}. 

\begin{table*}
\centering
\label{tab.M31_campaign}
\begin{minipage}{180mm}
\caption{Principal M31 classical nova surveys}
\begin{tabular}{lllllll}
\hline
Author(s)$/$Project      & Epoch      & Filter(s)        & Detector & Novae & Annual rate     & Reference(s) \\
\hline                                                    
Hubble                 & 1909--1927 & B                  & Plates   &  85   & $\sim 30$       & \citet{1929ApJ....69..103H}       \\
Arp                    & 1953--1954 & B                  & Plates   &  30   & $26 \pm 4$      & \citet{1956AJ.....61...15A}       \\
Rosino {\it et al.}    & 1955--1986 & B                  & Plates   & 142   & -               & \citet{1964AnAp...27..498R, 1973A+AS....9..347R,1989AJ.....97...83R} \\
Ciardullo {\it et al.} & 1982--1986 & B, H$\alpha$       & CCD      &  40   & -               & \citet{1987ApJ...318..520C, 1990ApJ...356..472C}    \\
Sharov \& Alksins      & 1969--1989 & B                  & Plates   &  21   & -               & \citet{1991Ap+SS.180..273S}       \\
Tomaney \& Shafter     & 1987--1989 & H$\alpha$          & CCD      &   9   & -               & \citet{1992ApJS...81..683T}       \\
Shafter \& Irby        & 1990--1997 & H$\alpha$          & CCD      &  72   & $37_{-8}^{+12}$  & \citet{2001ApJ...563..749S}       \\
Rector {\it et al.}    & 1995--1999 & H$\alpha$          & CCD      &  44   & -               & \citet{1999AAS...195.3608R}       \\
AGAPE                  & 1994--1996 & $R,I$              & CCD      &  12   & -               & \citet{2004A+A...421..509A}       \\
POINT-AGAPE            & 1999--2002 & $r'$, $i'$, $g'$   & CCD      &  20   & $65_{-15}^{+16}$ & \citet{2006MNRAS.369..257D}       \\
NMS                    & 2001--2002 & $R,I$              & CCD      &   2   & -               & \citet{2004A+A...415..471J}       \\
WeCAPP                 & 1997--2008 & $R,I$              & CCD      &  91   & -               & This work \\
\hline
\end{tabular}
\end{minipage}
\end{table*}

A list of all the campaigns with published novae in M31 that we are aware of is shown in Table~\ref{tab.M31_campaign}, 
with most of the data compiled by \cite{2001ApJ...563..749S} and \cite{2004MNRAS.353..571D}.

Despite the extensive search towards M31, most of previous studies have only sparse observations 
and thus make the analysis of nova light curve rather difficult. Our WeCAPP project is dedicated 
to monitoring M31 with up to. 4 we categorize our nova candidates according to the 
classification scheme of \cite{2010AJ....140...34S}. We apply the power-law decline proposed 
by ~\cite{2006ApJS..167...59H} to fit the smooth class light curves in Sect. 4.1. Novae showing 
cusp, oscillation or jitter features in their light curves are presented in Sect. 4.2 - 4.4. 
We then correlate our nova candidates with literature to search for recurrent novae in Sect. 5 
We show the rate of decline of your nova candidates and the distribution
of their speed class in Sect. 6 and end the paper with the conclusions in Sect.
7. All the light curves in our catalogue are presented in the Appendix.

\section{Observations and data reduction}

\begin{figure}
  \centering
  \includegraphics[width=0.45\textwidth]{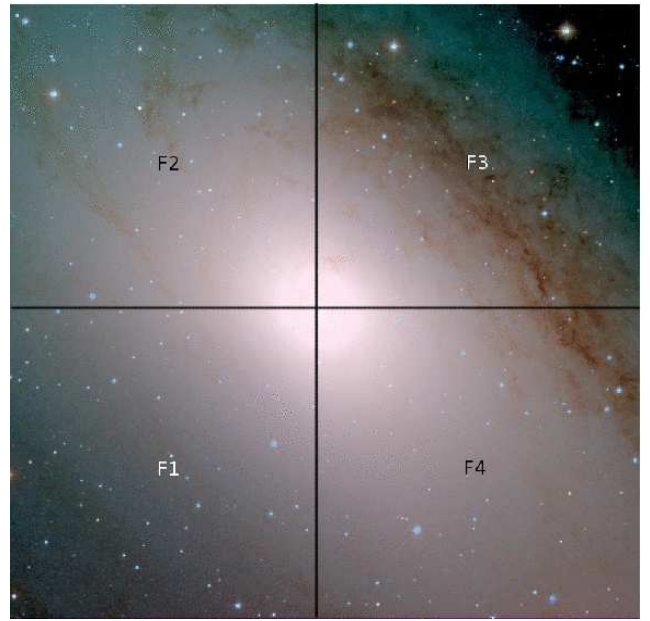}
  \caption{M31 composite image ($V$-, $R$-, and $I$-band) taken 
      at the Calar Alto Observatory. The black lines mark the four pointings 
      (F1 to F4) from the Wendelstein Observatory to mosaic the full FOV of 
      the Calar Alto Observatory.}
  \label{fig.m31_farbig}
\end{figure}

The WeCAPP project \citep{2001A+A...379..362R} was a dedicated survey
to search for microlensing events towards our neighboring galaxy
M31. We continuously monitored the bulge of M31 (when it was visible,
when the weather was cooperative and when there was an observer)
between September 1997 and March 2008 using the 0.8 m telescope of the
Wendelstein Observatory located in the Bavarian Alps. 
The data was taken optimally on a daily basis in both \textit{R} and
\textit{I} filters with a field of view of 8$\farcm$3 $\times$ 8$\farcm$3. From
June 1999 to February 2002 we further extended our observations with
the 1.23 m (17$\farcm$2 $\times$ 17$\farcm$2 FOV) telescope of the Calar Alto
Observatory in Spain. After 2002 we use the Wendelstein telescope
solely to mosaic the full Calar Alto field of view with four pointings.
The position of these four pointing are indicated in Fig. \ref{fig.m31_farbig}.

The data volume and quality of the four pointings (F1, F2,
F3, F4) drastically differs during the 11 seasons.
A list of the number of nights observed in each season is shown in 
Table \ref{tab.nights}.
A detailed overview of the observaions can be found in 
\cite{2006A+A...445..423F} and Riffeser et. al. (in prep.).

\begin{table}[ht]
  \setlength{\tabcolsep}{1.5mm}
  \begin{center}
  \caption{Lists of the analyzed nights per season from the 11-year WeCAPP campaign}
  \begin{tabular}{c|rrrr|rrrr}
    \hline\hline
     season        &        \multicolumn{4}{c|}{$R$-band}         &        \multicolumn{4}{c}{$I$-band}      \\
    \hline
                   &       F1 &       F2 &       F3 &       F4 &    F1 &       F2 &       F3 &       F4 \\ 
    \hline\hline
     1997 - 1998 &       36 &        7 &        1 &        4 &    33 &        7 &        0 &        3 \\
     1998 - 1999 &       33 &        1 &        1 &        1 &    28 &        1 &        1 &        1 \\
     1999 - 2000 &      154 &        0 &        0 &        0 &   145 &        0 &        0 &        0 \\
     2000 - 2001 &      184 &      108 &      124 &      108 &   159 &       89 &      104 &       89 \\
     2001 - 2002 &      240 &      136 &      159 &      136 &   212 &      119 &      140 &      119 \\
     2002 - 2003 &       34 &       18 &       24 &       18 &    30 &       16 &       24 &       18 \\
     2003 - 2004 &       35 &       24 &       29 &       31 &    33 &       21 &       26 &       29 \\
     2004 - 2005 &       25 &       23 &       26 &       25 &    19 &       16 &       19 &       19 \\
     2005 - 2006 &       30 &       26 &       28 &       29 &    26 &       20 &       22 &       23 \\
     2006 - 2007 &      107 &      106 &      103 &      103 &    48 &       45 &       46 &       47 \\
     2007 - 2008 &       62 &       56 &       52 &       58 &    36 &       35 &       35 &       38 \\
    \hline
     total         &      940 &      505 &      547 &      513 &   769 &      369 &      417 &      386 \\
    \hline\hline
  \end{tabular}
  \tablefoot{Each season 
        starts from the 1st of May until the 30th of April in the next year. The WeCAPP campaigns began 
        in 1997 focusing on F1 with the Wendelstein Observatory. From 1999 until 2002 we extended our
        observations by including the Calar Alto Observatory, which boosted the number of images taken in 
        these seasons. From 2002 on, 
        we use the Wendelstein Observatory solely and mosaic the full Calar Alto FOV with four pointings.}
  \label{tab.nights}
  \end{center}
\end{table}

\begin{figure}[ht]
    \centering
    \includegraphics[width=0.43\textwidth]{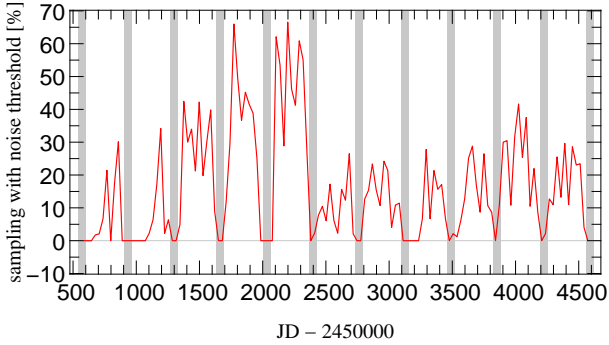}
    \caption{Fraction of good quality data points in $t$ averaged over the survey area. The definition
             of good quality is given in the text. The vertical grey zones indicate the time when M31 is not
             observable from the location of the telescopes during May and June). 
             0\% indicates we have no observations during the night.}
    \label{fig.lownoise_sampling}
  \end{figure}

  To quantify a realistic time sampling of the survey we define ``good quality 
  data points'' as data points with PSF fluxes with an error below $0.4
  \times 10^{-5}\mathrm{Jy}$. In Fig.~\ref{fig.lownoise_sampling} we show
  for every night the fractional area of pixels with errors below this limit. 
  0\% indicates we have no observations during the night.
  
  \begin{figure}[ht]
    \centering
    \includegraphics[width=0.4\textwidth]{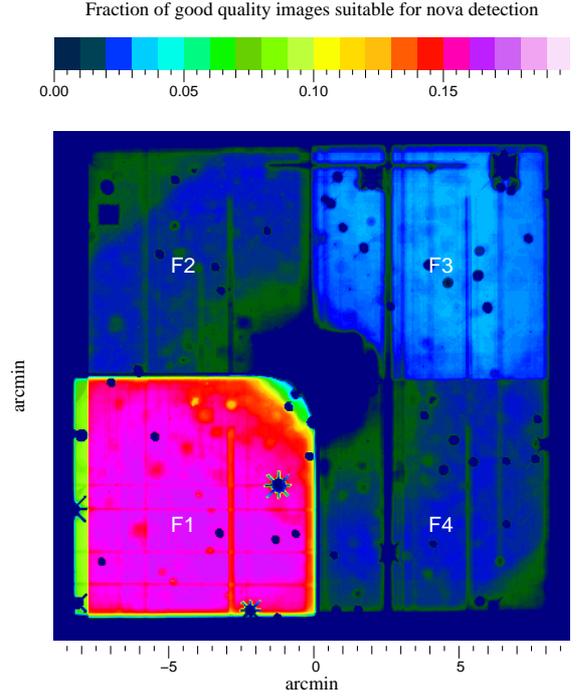}
    \caption{Fraction of good quality data points in $(x,y)$ averaged over time $t$.
             The definition of good quality is given in the text.
             The low fraction in the central part is caused by the high noise of M31 itself.}
    \label{fig.lownoise}
  \end{figure}
  
  Fig.~\ref{fig.lownoise} shows the spatial variation of the fraction of all data with 
  flux errors below the flux error limit averaged over 11 seasons. It demonstrates that we expect most of our novae in field F1 
  and fewer in the fields F2, F3, and F4. The field F1 was observed much more frequently than the 
  other one because it is the subfield with highest lensing probability.

The data was then reduced by our customized pipeline MUPIPE \citep[see ][]{2002A+A...381.1095G}, 
which performs CCD reduction, position alignment, photometric alignment, frame stacking and difference 
imaging following the algorithm of \cite{1998ApJ...503..325A}.

After the difference imaging, we perform PSF photometry on each pixel as follows. First, we extracted
the PSF from several isolated, bright and unsaturated reference stars. Then we fit this PSF to all 
variable sources. Finally, we integrate the count rates over the area of the PSF to determine the flux of the source.

The results of the project are presented in 
\cite{2003ApJ...599L..17R, 2008ApJ...684.1093R} and partially contributed to \cite{2010ApJ...717..987C}. 
In addition to the original 
microlensing targets, the intensive observations in two bands also yields more than 20,000 variables 
in the bulge of M31 \citep{2006A+A...445..423F} and the nova candidates presented in this paper.

\section{Nova detection}

To establish an automatic detection for nova candidates, we apply the following criteria for candidates selection 
based on the measured $R$-band PSF flux (as mentioned in Sect. 2):\\

\begin{itemize}
\item The significance for variability must be 10$\sigma$ relative to the baseline and the measured flux excess of the variable source must be a local maximum around neighbouring pixels at a given time step. 
Note that $\sigma$ throughout this paper refers to the errors of the individual PSF flux excess measurements.
\item The variable source must have a measured flux excess larger than $4\times10^{-5}$ Jy in \textit{R}-band (corresponding to 
$m_R$ = $-$2.5 log($\frac{4\times 10^{-5} Jy}{F_{\mathrm{Vega,}R}}$) $\sim$ 19.7 mag, with $F_{\mathrm{Vega,}R}$ = 3060 Jy being the flux 
of Vega in the $R$-band) and the first measurement after the measured maximum flux excess must have a flux 
excess $>~2\times10^{-5}$ Jy.
\item To use the eruptive nature of novae, we define the strength $s$ of the outburst:
\begin{equation}
  s = \frac{\Delta F_{\max}/\sigma _{\max} ^2+\Delta F _{\max+1}/\sigma _{\max+1} ^2}{1/\sigma _{\max} ^2+1/\sigma _{\max+1} ^2}
\end{equation}
where $\Delta F_{\max}$ is the measured maximum flux excess relative to the reference image and $\Delta F_{\max+1} $ is the first measurement after 
the measured maximum flux excess. The $\sigma _{\max} $ and $\sigma _{\max+1}$ are the errors in the measurements of the flux excess. We require 
$s$ $>$ 4.6 $\times$ 10$^{-5}$ Jy nova detection.
\begin{table}[!h]
\centering
\caption{Detection criteria for nova candidates}
\begin{tabular}{lr}
\hline
Criterion       & Number \\
\hline
Full light curves & 4043256           \\
Local flux maximum with $s>$4.6$\times$10$^{^{-5}}$Jy and $a>$4.7 &  1005 \\
Masking of bright stars & 156 \\
Grouping & 105 \\
Inspection by eye & 91 \\
\hline
\label{tab.detect}
\end{tabular}
\end{table}
\item To avoid false contamination from periodically varying sources, we define the asymmetry $a$ between positive and negative 
outliers in the light curve relative to the baseline:
\begin{equation}
  a=\frac{\mathrm{Number~of~data~points~with~}\Delta F>5\sigma}{\mathrm{Number~of~data~points~with~}\Delta F<-5\sigma} - 1.
\end{equation}
This quantity $a$ is useful in filtering out normal variable sources, which have $a$ $\sim$ 0, while the eruptive nature of 
novae lead to $a~\gg$ 1. We empirically require $a$ to be larger than 4.7 to be suitable for nova detection.
\item We than apply a special mask to filter false detections around bright stars, especially spikes.\\
\item After the masking, we apply a group algorithm to find multiple pixel detections connecting to the same nova candidate in different time steps.\\
\item In the last step, we inspect the difference images and light curves by eye to make sure that no image
artefact escapes our detection and is misinterpreted as a nova.
\end{itemize}
We combine the criteria 1-4 into one single step. The detections filtered out by each steps are shown in Table. \ref{tab.detect}. 

Among the nova candidates, 24 are discovered by WeCAPP project for the first time, 
while 5 of them are known but were not officially published and can be
found on the CBAT\footnote{M31 (Apparent) Novae Page, http://www.cfa.harvard.edu/iau/CBAT\_M31.html} or 
Extragalactic Novae\footnote{www.rochesterastronomy.org/novae.html} webpages. The rest of the nova candidates 
are published and can be found in the literature, 
see e.g. \cite{2007A+A...465..375P, 2010AN....331..187P}\footnote{An up-to-date online-version of the 
catalog can be found at http://www.mpe.mpg.de/$\sim$m31novae/opt/m31/index.php}. The positions and light curves of these 91 novae
are presented in Table \ref{tab.cat} and in the Appendix.

\section{Nova taxonomy}
Although all novae slightly differ, it is possible to group novae by their 
light-curve or spectroscopic properties. One of the commonly used methods to characterize novae 
is the `speed class' proposed by \cite{1964gano.book.....P}, who categorized novae according to 
their light-curve evolution and described the decline time-scale by the time needed to drop by
2 magnitudes below the maximum ($t_2$). 
\cite{1992AJ....104..725W}
did a thorough study of the spectroscopic properties of the novae, and categorized novae into 
Fe (galactic thick disk novae) or He (galactic disk novae) group according to the most prominent features in their 
spectra. 
Della Valle \& Livio (1998) further established the connection between the 
speed class and spectroscopic classification. They found 
that fast novae are mainly related to the He novae, while the slow novae tend to show Fe II features in their 
spectra. The prospoed 
explanation behind is that He novae are from the galactic disk and prone to have massive white dwarfs, 
thus having fast and steep decline. On the other hand, the Fe II novae originate from the less
massive population II stars in the galactic thick disk, and hence have a slow decline. 

The speed class is not enough to fully account for the differences between novae.
\cite{2010AJ....140...34S} gathered 93 galactic novae from the  
American Association of Variable Star Observers (AAVSO) and made a 
thorough study using the complete coverage of their light curves. 

They suggested to classify the
novae according to their distinct features during their decline, such as the plateau, the cusp by the 
secondary brightening and the dip by the dust.

In this section we classify our nova candidates (if possible) following the taxonomy proposed 
by \cite{2010AJ....140...34S}. Readers are referred to Table 3 and Figure 2 
in \cite{2010AJ....140...34S} for the definition and exemplary light curves for different nova classes. 
Note that the classification scheme of \cite{2010AJ....140...34S} is based on the $V$-band magnitude, 
while we are using $R$-band and might be affected by the strong H$\alpha$ emission. We thus check  
our $I$-band light curve, which does not affected by the strong H$\alpha$ emission, and identify  
the apparent features in the nova classification scheme of \cite{2010AJ....140...34S} in both $R$ and $I$-band.

\subsection{S Class and the universal decline law}

The S-class novae have smooth light curves following the universal power-law decline 
($F \propto t^{-1.75}$) due to free-free emission expanding shell as proposed by 
~\cite{2006ApJS..167...59H}. In principle, the classification scheme of 
\cite{2010AJ....140...34S} is based on the fact that all the light curves originate from the S-class. 
The S-class is indeed consistent to the vast majority of our nova candidates.
To verify the universal decline law, we thus fit our candidate 
light curves with a 4-parameter formula: 

\begin{equation}
\Delta F = f_b + f_0 \times (t-t_0)^{\alpha},
\label{eq.free_t0}
\end{equation}

where $f_b$ is the baseline level and will be different from zero in cases where the nova candidate flux is present in the 
reference frame used in difference imaging or there is a variable close to it (see e.g. the light curve of WeCAPP-N10 in 
the appendix). $f_0$ gives the proportional factor between the 
flux and time, $t_0$ is the onset of nova outburst and $\alpha$ is the index of the power-law decline. 
After the first iteration, we found that some candidates have unreasonable $t_0$ long before the nova eruption. 
For such events, we use a 5-parameter formula

\begin{equation}
\Delta F = f_b + f_0 \times (t-t_0)^{\alpha},\quad t_0 \equiv t_{_{-1}} + \delta^2
\label{eq.fixed_t0}
\end{equation}

with $t_{_{-1}}$ fixed at the last data point in the baseline just before the eruption to avoid unreasonable $t_0$.
The best-fit parameters for equations (\ref{eq.free_t0}) and (\ref{eq.fixed_t0}) are given in Table 
\ref{tab.free_t0}. 

For the S-class nova, we first tried to fit the power-law decline for all the nova candidates. A candidate is classified as S-class nova only
when the fitting routine finds a solution for either equation \ref{eq.free_t0}
or equation \ref{eq.fixed_t0}. N01, N09, N17, N24, N41, N49, N54, N58 and N77 are not attributed to
S-class because the fitting routine failed to find a solution.

Our best-fit value of $\alpha$ from Table \ref{tab.free_t0} for free $t_0$ solely and combined 
with fixed $t_0$ are $-$1.51 and $-$1.32, respectively. 

\begin{figure*}
  \centering
  \includegraphics[scale=0.95]{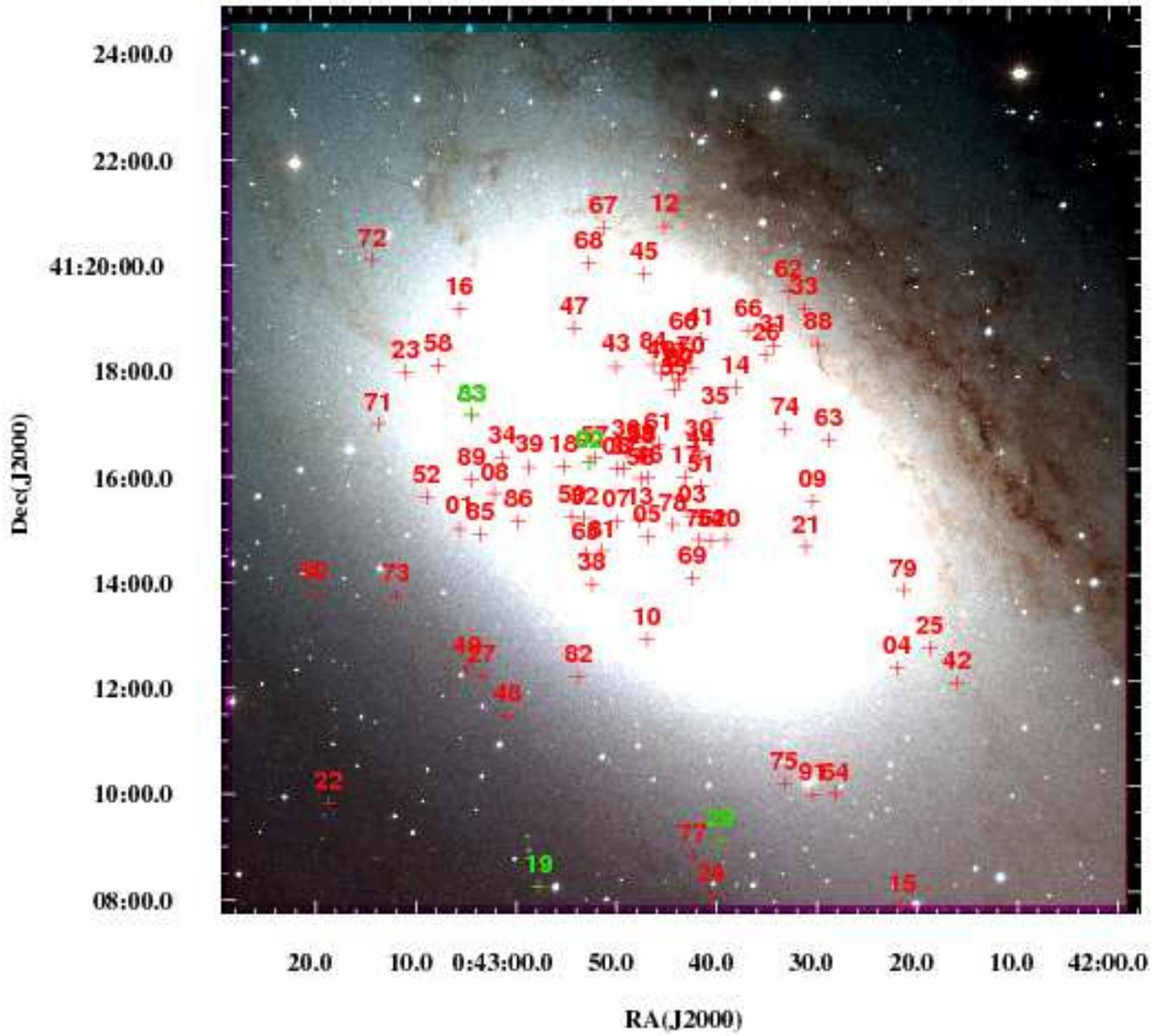}
  \caption{Distribution of the WeCAPP nova candidates. 
    The recurrent nova candidates (see Sect. 5) are marked in green. 
    The overlaying image is a three-color-combined image
    using the observations obtained from Calar Alto observatories in $V$, $R$ and $I$-band. The image has 
    a size of 17$\farcm$2 $\times$ 17$\farcm$2.}
  \label{fig.all}
\end{figure*}

\clearpage

\begin{table*}[!h]
\centering
\begin{sideways}
\begin{minipage}{275mm}
\caption{WeCAPP nova catalogue}
\begin{tabular}{|ccccrlllll|}
\hline
Name & RA(2000)     & Dec(2000)    & $t_{\max}$ & $\Delta t_{\max}$ & Class & CBAT & Discovery (and light curve) reference(s)& X-ray obs. & Spectroscopic obs. \\
 & [h:m:s] & [d:m:s] & & [day] & & & & &  \\
\hline
N01 & 00:43:05.37 & 41:14:59.2 &  745.52 &  29.01 & Unclassified & 1997-10e & 1997-14 in \cite{2001ApJ...563..749S}               &                                                                           &                                   \\ 
N02 & 00:42:52.35 & 41:16:13.2 &  750.45 &  33.95 & Unclassified & 1997-10f & 1997-10 in \cite{2001ApJ...563..749S}               &                                                                           &                                   \\
N03 & 00:42:42.13 & 41:15:10.4 &  753.55 &  37.05 & Unclassified & 1997-11a & 1997-07 in \cite{2001ApJ...563..749S}               & \cite{2007A+A...465..375P}                                                &                                   \\ 
\multicolumn{8}{|c}{}                                                                                                             & \cite{2010A+A...523A..89H,2010arXiv1010.1461H}                            &                                   \\
N04 & 00:42:21.76 & 41:12:16.2 &  753.55 &  37.05 & Unclassified & 1997-10c & 1997-02 in \cite{2001ApJ...563..749S}               &                                                                           &                                   \\
N05 & 00:42:46.64 & 41:14:49.2 & 1109.48 & 243.19 & Unclassified & 1998-09d & IAUC 7015, \cite{2000AstL...26..433S}               &                                                                           & Fe II, \cite{2011arXiv1104.0222S} \\
N06 & 00:42:49.65 & 41:16:06.5 & 1251.30 &   2.00 & Unclassified & 1999-02a & This work                                     &                                                                           &                                   \\ 
N07 & 00:42:49.69 & 41:15:05.6 & 1359.55 &   0.93 & Cusp         & 1999-06a & IAUC 7218, PAV-78668 in \cite{2004MNRAS.351.1071A}  &                                                                           & Fe II, \cite{2011arXiv1104.0222S} \\
N08 & 00:43:01.85 & 41:15:38.4 & 1372.62 &   1.01 & Smooth       & 1999-06b & \cite{1999BAAS...31.1420R}                          &                                                                           &                                   \\
N09 & 00:42:30.11 & 41:15:27.3 & 1719.62 & 553.15 & Unclassified & 2000-06a & This work                                     &                                                                           &                                   \\
N10 & 00:42:46.75 & 41:12:51.9 & 1726.63 &   1.00 & Cusp         & 2000-08b & \cite{2007A+A...465..375P}                          &                                                                           &                                   \\
N11 & 00:42:43.97 & 41:17:55.5 & 1754.64 &  18.01 & Oscillation  & 2000-07a & PACN-00-01 in \cite{2004MNRAS.353..571D}            & \cite{2005A+A...442..879P,2007A+A...465..375P}                            &                                   \\
\multicolumn{8}{|c}{}                                                                                                             & \cite{2006ApJ...643..844O}                                                &                                   \\
N12 & 00:42:44.65 & 41:20:40.6 & 1755.65 &   1.02 & Cusp         & 2000-07b & PACN-00-03 in \cite{2004MNRAS.353..571D}            &                                                                           &                                   \\ 
N13 & 00:42:47.45 & 41:15:07.8 & 1763.66 &   1.02 & Cusp         & 2000-08a & \cite{2007A+A...465..375P}                          & \cite{2005A+A...442..879P,2007A+A...465..375P}                            &                                   \\ 
N14 & 00:42:37.70 & 41:17:37.8 & 1766.64 &   1.00 & Cusp         & 2000-08d & PACN-00-04 in \cite{2004MNRAS.353..571D}            &                                                                           &                                   \\ 
N15 & 00:42:21.49 & 41:07:47.3 & 1932.34 &   1.04 & Cusp         & 2001-01a & This work                                     & \cite{2010A+A...523A..89H,2010arXiv1010.1461H}                            &                                   \\
N16 & 00:43:05.26 & 41:19:08.2 & 1948.34 &   4.02 & Unclassified & 2001-01b & This work                                     &                                                                           &                                   \\
N17 & 00:42:42.82 & 41:15:55.2 & 1940.33 &   6.04 & Unclassified & 2001-01c & This work                                     &                                                                           &                                   \\
N18 & 00:42:54.95 & 41:16:09.2 & 1948.34 &   4.02 & Unclassified & 2001-02a & This work                                     &                                                                           &                                   \\
N19 & 00:42:57.75 & 41:08:12.3 & 2097.56 & 124.25 & Unclassified & 2001-07b & This work                                     &                                                                           &                                   \\ 
N20 & 00:42:38.76 & 41:14:44.4 & 2097.56 & 124.25 & Unclassified & 2001-07c & This work                                     &                                                                           &                                   \\
N21 & 00:42:30.79 & 41:14:36.1 & 2130.63 &   3.00 & Unclassified & 2001-07d & IAUC 7674, PACN-01-01 in \cite{2004MNRAS.353..571D} &                                                                           &                                   \\
N22 & 00:43:18.62 & 41:09:49.0 & 2094.56 & 121.25 & Smooth       & 2001-07a & PAV-74935 in \cite{2004MNRAS.351.1071A}             & \cite{2005A+A...442..879P,2007A+A...465..375P}                            &                                   \\
N23 & 00:43:10.62 & 41:17:58.0 & 2163.65 &  28.02 & Unclassified & 2001-08b & PACN-01-03 in \cite{2004MNRAS.353..571D}            &                                                                           &                                   \\
N24 & 00:42:40.60 & 41:07:59.9 & 2163.65 &  28.02 & Unclassified & 2001-08c & PACN-01-04 in \cite{2004MNRAS.353..571D}            &                                                                           &                                   \\
N25 & 00:42:18.52 & 41:12:39.3 & 2163.65 &  28.02 & Unclassified & 2001-08a & IAUC 7684, PACN-01-02 in \cite{2004MNRAS.353..571D} &                                                                           &                                   \\
N26 & 00:42:34.62 & 41:18:13.0 & 2151.60 &   0.99 & Smooth       & 2001-08d & IAUC 7709, PAC-26277 in \cite{2004MNRAS.351.1071A}  & \cite{2005A+A...442..879P,2007A+A...465..375P}                            &                                   \\
N27 & 00:43:03.31 & 41:12:11.5 & 2190.48 &   1.90 & Jitter       & 2001-10a & IAUC 7729, PACN-01-06 in \cite{2004MNRAS.353..571D} & \cite{2007A+A...465..375P}                                                & Fe II, \cite{2011arXiv1104.0222S} \\
\multicolumn{7}{|c}{}                                                       & NMS2 in \cite{2004A+A...415..471J}                  & \cite{2010A+A...523A..89H,2010arXiv1010.1461H}                            &                                   \\
N28 & 00:42:47.21 & 41:16:18.7 & 2197.32 &   0.98 & Oscillation  & 2001-10c & This work                                     &                                                                           &                                   \\
N29 & 00:42:39.59 & 41:09:02.9 & 2299.32 &   2.99 & Unclassified & 2001-12b & This work                                     &                                                                           &                                   \\
N30 & 00:42:41.44 & 41:16:24.6 & 2266.30 &   0.92 & Smooth       & 2001-12a & IAUC 7794                                           &                                                                           & Fe II, \cite{2011arXiv1104.0222S} \\
\hline
\end{tabular}
\label{tab.cat}
\end{minipage}
\end{sideways}
\end{table*}
\addtocounter{table}{-1}
\begin{table*}[!h]
\centering
\begin{sideways}
\begin{minipage}{275mm}
\caption{WeCAPP nova catalogue continued.}
\begin{tabular}{|ccccrlllll|}
\hline
Name & RA(2000)     & Dec(2000)    & $t_{\max}$ & $\Delta t_{\max}$ & Class & CBAT & Discovery (and light curve) reference(s)& X-ray obs. & Spectroscopic obs. \\
 & [h:m:s] & [d:m:s] & & [day] & & & & &  \\
\hline
N31 & 00:42:33.89 & 41:18:24.0 & 2282.31 &   6.02 & Smooth       & 2002-01b & IAUC 7794, PAV-26285 in \cite{2004MNRAS.351.1071A}  & \cite{2005A+A...442..879P,2007A+A...465..375P}                            & He/N, \cite{2011arXiv1104.0222S}  \\
N32 & 00:42:52.89 & 41:15:10.4 & 2283.30 &   0.99 & Cusp         & 2002-01a & IAUC 7794, PAV-79136 in \cite{2004MNRAS.351.1071A}  &                                                                           &                                   \\
N33 & 00:42:30.74 & 41:19:05.9 & 2325.38 &   4.02 & Smooth       & 2002-02a & This work                                     &                                                                           &                                   \\
N34 & 00:43:01.08 & 41:16:19.9 & 2476.54 &  13.00 & Smooth       & 2002-07a & IAUC 7937, IAUC 7938                                &                                                                           &                                   \\
N35 & 00:42:39.74 & 41:17:03.3 & 2521.57 &  38.01 & Doubtful     & 2002-07b & This work                                     &                                                                           &                                   \\
N36 & 00:42:48.66 & 41:16:26.3 & 2573.63 &  26.07 & Doubtful     & 2002-08b & This work                                     &                                                                           &                                   \\
N37 & 00:42:48.90 & 41:16:05.3 & 2661.25 &   6.82 & Smooth       & 2003-01b & This work                                     &                                                                           &                                   \\
N38 & 00:42:52.24 & 41:13:54.5 & 2797.53 &  91.24 & Doubtful     & 2003-01c & IAUC 8155                                           &                                                                           &                                   \\
N39 & 00:42:58.38 & 41:16:08.3 & 2797.53 &  91.24 & Smooth       & 2003-06a & IAUC 8155                                           &                                                                           &                                   \\
N40 & 00:42:45.12 & 41:17:54.0 & 2820.50 &  21.94 & Smooth       & 2003-06c & IAUC 8165                                           &                                                                           &                                   \\
N41 & 00:42:41.14 & 41:18:32.4 & 2832.56 &  12.05 & Unclassified & 2003-06d & IAUC 8165                                           &                                                                           &                                   \\
N42 & 00:42:15.85 & 41:11:59.9 & 2834.44 &   5.96 & Smooth       & 2003-07b & IAUC 8165, N3 in \cite{2005ASPC..330..449S}         &                                                                           &                                   \\
N43 & 00:42:49.64 & 41:18:02.0 & 2867.53 &   6.11 & Doubtful     & 2003-08a & IAUC 8210                                           &                                                                           &                                   \\
N44 & 00:42:41.20 & 41:16:16.0 & 2925.46 &  16.95 & Unclassified & 2003-08c & IAUC 8226                                           & \cite{2010arXiv1010.1461H}                                                &                                   \\ 
N45 & 00:42:46.74 & 41:19:47.4 & 2931.30 &  22.96 & Smooth       & 2003-09b & IAUC 8222, N5 in \cite{2005ASPC..330..449S}         &                                                                           &                                   \\
N46 & 00:42:46.45 & 41:15:55.6 & 2925.42 &  26.88 & Unclassified & 2003-10a & This work                                     &                                                                           &                                   \\
N47 & 00:42:53.78 & 41:18:46.2 & 2949.54 &   8.12 & Unclassified & 2003-11a & IAUC 8248                                           & \cite{2007A+A...465..375P}                                                &                                   \\ 
\multicolumn{8}{|c}{}                                                                                                             & \cite{2010A+A...523A..89H}                                               &                                   \\
N48 & 00:43:00.76 & 41:11:26.9 & 2978.22 &   6.86 & Smooth       & 2003-11b & IAUC 8253                                           & \cite{2007A+A...465..375P}                                                &                                   \\
\multicolumn{8}{|c}{}                                                                                                             & \cite{2010A+A...523A..89H}                                               &                                   \\
N49 & 00:43:04.73 & 41:12:21.9 & 2992.32 &   7.02 & Unclassified & 2003-12a & IAUC 8262, N8 in \cite{2005ASPC..330..449S}         & \cite{2007A+A...465..375P}                                                &                                   \\
N50 & 00:42:54.14 & 41:15:12.2 & 2994.32 &   2.00 & Smooth       & 2003-12b & IAUC 8262                           & \cite{2007A+A...465..375P}                                               &                                   \\ 
N51 & 00:42:41.18 & 41:15:45.0 & 3006.24 &  52.88 & Unclassified & 2004-01b & This work                     & \cite{2007A+A...465..375P}                                               &                                   \\ 
\multicolumn{8}{|c}{}                                                                                             & \cite{2010arXiv1010.1461H}                                               &                                   \\
N52 & 00:43:08.65 & 41:15:35.3 & 3039.29 &  38.95 & Smooth       & 2004-01a & N9 in \cite{2005ASPC..330..449S}    & \cite{2007A+A...465..375P}                                               &                                   \\ 
N53 & 00:42:47.28 & 41:16:21.4 & 3039.29 &  52.81 & Cusp         & 2004-02a & This work                     & \cite{2007A+A...465..375P}                                               &                                   \\  
N54 & 00:42:40.28 & 41:14:42.5 & 3254.44 & 193.14 & Unclassified & 2004-09a & IAUC 8404                           & \cite{2007A+A...465..375P}                                               & Fe II, \cite{2011arXiv1104.0222S} \\  
N55 & 00:42:43.90 & 41:17:35.0 & 3319.60 &   9.30 & Doubtful     & 2004-07a & \cite{2007A+A...465..375P}          & \cite{2007A+A...465..375P}                                               &                                   \\  
N56 & 00:42:47.24 & 41:15:54.5 & 3291.45 &   7.92 & Unclassified & 2004-10b & \cite{2007A+A...465..375P}          & \cite{2007A+A...465..375P}                                               &                                   \\ 
N57 & 00:42:51.84 & 41:16:18.2 & 3291.37 &   7.84 & Unclassified & 2004-10a & ATEL 346                            & \cite{2007A+A...465..375P}                                               &                                   \\ 
N58 & 00:43:07.46 & 41:18:04.6 & 3319.49 &  15.13 & Unclassified & 2004-11b & \cite{2007A+A...465..375P}          & \cite{2007A+A...465..375P}                                               & He/N, \cite{2011arXiv1104.0222S}  \\ 
\multicolumn{8}{|c}{}                                                                                             & \cite{2010A+A...523A..89H}                                               &                                   \\
N59 & 00:42:47.17 & 41:16:19.8 & 3319.49 &  15.13 & Smooth       & 2004-11f & CBAT                                & \cite{2007A+A...465..375P}                                               &                                   \\ 
N60 & 00:42:42.81 & 41:18:27.8 & 3319.60 &   9.30 & Unclassified & 2004-11a & \cite{2007A+A...465..375P}          & \cite{2007A+A...465..375P}                                               & Fe II, \cite{2011arXiv1104.0222S} \\  
\hline
\end{tabular}
\label{tab.cat1}
\end{minipage}
\end{sideways}
\end{table*}
\addtocounter{table}{-1}
\begin{table*}[!h]
\centering
\begin{sideways}
\begin{minipage}{275mm}
\caption{WeCAPP nova catalogue continued.}
\begin{tabular}{|ccccrlllll|}
\hline
Name & RA(2000)     & Dec(2000)    & $t_{\max}$ & $\Delta t_{\max}$ & Class & CBAT & Discovery (and light curve) reference(s)& X-ray obs. & Spectroscopic obs. \\
 & [h:m:s] & [d:m:s] & & [day] & & & & &  \\
\hline
N61 & 00:42:45.47 & 41:16:33.2 & 3348.42 &  28.93 & Unclassified & 2004-11d & \cite{2007A+A...465..375P}          & \cite{2007A+A...465..375P}                                               &                                   \\ 
N62 & 00:42:32.29 & 41:19:25.7 & 3346.36 &  25.92 & Cusp         & 2004-11c & CBAT                                & \cite{2007A+A...465..375P}                                               &                                   \\ 
N63 & 00:42:28.39 & 41:16:36.1 & 3382.36 &   4.09 & Unclassified & 2005-01a & \cite{2007A+A...465..375P}          & \cite{2007A+A...465..375P}                                               & Fe II, \cite{2011arXiv1104.0222S} \\ 
N64 & 00:42:28.10 & 41:09:54.7 & 3381.36 &  21.03 & Unclassified & 2004-12a & ATEL 379                            & \cite{2007A+A...465..375P}                                               &                                   \\ 
N65 & 00:42:52.79 & 41:14:28.8 & 3426.28 &  17.92 & Smooth       & 2005-02a & ATEL 421                            & \cite{2007A+A...465..375P}                                               &                                   \\
\multicolumn{8}{|c}{}                                                                                             & \cite{2010A+A...523A..89H,2010arXiv1010.1461H}                           &                                   \\ 
N66 & 00:42:36.37 & 41:18:41.8 & 3427.38 &   1.03 & Doubtful     & 2005-02b & This work                     &                                                                          &                                   \\ 
N67 & 00:42:50.80 & 41:20:39.8 & 3592.47 & 100.43 & Cusp         & 2005-07a & \cite{2007A+A...465..375P}          & \cite{2010A+A...523A..89H}                                               & Fe II, \cite{2011arXiv1104.0222S} \\ 
N68 & 00:42:52.25 & 41:19:59.4 & 3635.37 &  15.94 & Smooth       & 2005-09a & CBAT                                & \cite{2010A+A...523A..89H}                                               & Fe II, ATEL 850                   \\ 
N69 & 00:42:42.11 & 41:14:01.1 & 3635.59 &   9.30 & Unclassified & 2005-09d & This work                     & \cite{2010A+A...523A..89H}                                               &                                   \\ 
N70 & 00:42:42.12 & 41:18:00.3 & 3661.54 &   1.17 & Unclassified & 2005-10b & ATEL 651                            & \cite{2010A+A...523A..89H}                                               &                                   \\ 
                                                                                           \\
N71 & 00:43:13.42 & 41:16:58.9 & 3863.57 &  50.27 & Smooth       & 2006-04a & ATEL 805                            & \cite{2010A+A...523A..89H,2010AN....331..193H}                           &                                   \\ 
N72 & 00:43:13.93 & 41:20:05.5 & 3863.57 &  50.27 & Smooth       & 2006-05a & This work                     & \cite{2010A+A...523A..89H}                                               &                                   \\ 
N73 & 00:43:11.81 & 41:13:44.7 & 3880.53 &   2.98 & Unclassified & 2006-06a & This work                     & \cite{2010A+A...523A..89H}                                               & Fe II, ATEL 850                   \\ 
N74 & 00:42:32.77 & 41:16:49.1 & 3867.57 &  54.27 & Unclassified & 2006-06b & ATEL 829                            & \cite{2010A+A...523A..89H,2010arXiv1010.1461H}                           &                                   \\ 
N75 & 00:42:33.17 & 41:10:06.8 & 3984.41 &   3.89 & Smooth       & 2006-09a & \cite{2007A+A...469..115C}          & \cite{2010A+A...523A..89H}                                               &                                   \\ 
N76 & 00:42:41.45 & 41:14:44.5 & 4000.42 &   8.92 & Unclassified & 2006-09b & ATEL 884                            & \cite{2010A+A...523A..89H}                                               &                                   \\ 
N77 & 00:42:42.39 & 41:08:45.6 & 3999.40 &   4.98 & Unclassified & 2006-09c & ATEL 887, \cite{2011ApJ...727...50S}& \cite{2010A+A...523A..89H,2010arXiv1010.1461H}                           & Fe II, \cite{2011arXiv1104.0222S} \\ 
N78 & 00:42:44.05 & 41:15:02.1 & 4096.47 &   1.09 & Unclassified & 2006-11b$^{\dagger}$ & This work           &                                                                          &                                   \\ 
N79 & 00:42:21.08 & 41:13:45.4 & 4095.38 &   5.02 & Smooth       & 2006-12a & This work                     & \cite{2010A+A...523A..89H,2010arXiv1010.1461H}                           & Fe II, \cite{2011arXiv1104.0222S} \\
N80 & 00:42:43.22 & 41:17:48.4 & 4095.52 &   5.08 & Unclassified & 2006-12c & ATEL 973                            & \cite{2010A+A...523A..89H,2010arXiv1010.1461H}                           &                                   \\
N81 & 00:42:51.15 & 41:14:33.5 & 4122.43 &   6.14 & Unclassified & 2007-01a & CBAT                                & \cite{2010A+A...523A..89H,2010arXiv1010.1461H}                           &                                   \\
N82 & 00:42:53.61 & 41:12:09.9 & 4166.30 &  13.03 & Smooth       & 2007-03a & CBAT                                & \cite{2010A+A...523A..89H,2010arXiv1010.1461H}                           &                                   \\
N83 & 00:43:04.05 & 41:17:08.3 & 4296.48 &  24.96 & Smooth       & 2007-07a & ATEL 1131                           & \cite{2010arXiv1010.1461H}                                               &                                   \\
N84 & 00:42:45.91 & 41:18:04.4 & 4297.50 &   1.01 & Smooth       & 2007-07b & ATEL 1139                           & \cite{2010arXiv1010.1461H}                                               & Fe II, ATEL 1186                  \\
N85 & 00:43:03.29 & 41:14:53.0 & 4307.48 &   9.02 & Smooth       & 2007-07c & ATEL 1146                           & \cite{2010arXiv1010.1461H}                                               & Hybrid or He/N, ATEL 1186         \\
N86 & 00:42:59.49 & 41:15:06.6 & 4337.45 &   2.95 & Unclassified & 2007-07d & ATEL 1162                           & \cite{2010arXiv1010.1461H}                                               &                                   \\
N87 & 00:42:43.30 & 41:17:44.1 & 4314.55 &  15.01 & Smooth       & 2007-07e & ATEL 1156                           & \cite{2010arXiv1010.1461H}                                               & Fe II, ATEL 1186                  \\
N88 & 00:42:29.39 & 41:18:24.8 & 4356.49 &  16.92 & Smooth       & 2007-08c & IAUC 7664, ATEL 1198                & \cite{2010arXiv1010.1461H}                                               &                                   \\
N89 & 00:43:04.18 & 41:15:54.1 & 4425.30 &  14.81 & Unclassified & 2007-11c & ATEL 1275                           & \cite{2010arXiv1010.1461H}                                               & Fe II, \cite{2011arXiv1104.0222S} \\
N90 & 00:43:19.98 & 41:13:46.3 & 4444.61 &  18.35 & Smooth       & 2007-12b & This work                     & ATEL 1360, ATEL 1647                                                     & He/N, \cite{2011arXiv1104.0222S}  \\
\multicolumn{8}{|c}{}                                                                                             & \cite{2009ApJ...705.1056B}                                               &                                   \\
N91 & 00:42:30.37 & 41:09:53.6 & 4508.27 &  0.98  & Unclassified & 2008-02a & ATEL 1380                           & \cite{2010arXiv1010.1461H}                                               &                                   \\ 
\hline
\end{tabular}
\label{tab.cat1}
\tablefoot{We show the position and the time of maximum flux (expressed in JD $-$ 2450000) of the nova candidates in columns 2, 3 and 4. The uncertainty of the position is smaller than 0$\farcs$1 (see Riffeser et. al., in prep.). The uncertainty in the time of maximum flux $\Delta t_{\max}$ in column 5 is derived from the time difference between $t_{\max}$ and the last measurement before $t_{\max}$. The light curve classification is shown in column 6, with 'unclassified' indicates those novae we are not able to classify and 'doubtful' indicates the novae are more similar to other variables than novae. In column 7 we give the corresponding CBAT nomenclature. Column 8 list the references for discovery (and light curves) in optical. The spectroscopic and X-ray observations are shown in column 9 and 10. The 24 novae without discovery references are newly discovered by WeCAPP. $\dagger$ We also detected M31N-2006-12d on the same position, which is possibly rebrightening of M31N-2006-11b given the short time difference.}
\end{minipage}
\end{sideways}
\end{table*}
\clearpage

\noindent
The power-law index for a free $t_0$ is close to the 
value given by ~\cite{2006ApJS..167...59H}, while the value of $\alpha$ from a combination 
of both free and fixed $t_0$ deviates from $-$1.75, which indicates that we might have missed the 
true eruption date for some of the novae.
Note that we constrain the value of power-law index $\alpha$ from the $R$-band images, which are contaminated by 
the H$\alpha$ emission line and might differ from the universal power-law index from ~\cite{2006ApJS..167...59H}. 

\begin{figure}
  \centering
  \includegraphics[scale=0.57]{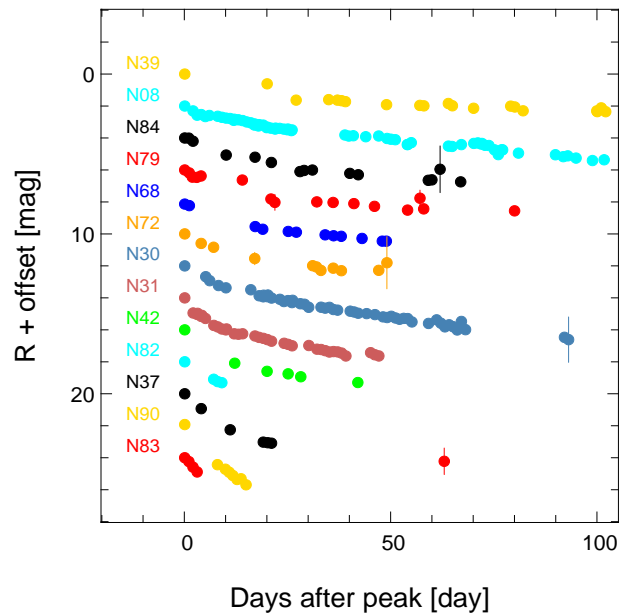}
  \caption{S Class novae with free $t_0$. The single offsets are $-$15.04 for N08, $-$3.59 for N30, 
$-$2.81 for N31, 2.84 for N37, $-$18.02 for N39, $-$0.88 for N42, $-$9.89 for N68, $-$8.06 for N72, 
$-$11.81 for N79 and 1.02 for N82, 5.14 for N83, $-$13.66 for N84 and 5.37 for N90 respectively. 
Note that for most of the  data points the error bars are smaller than the symbol of the data points. 
Here we only show the decline part of the light curve. 
Full light curves can be found in the appendix.}
  \label{fig.s-class_free_t0}
\end{figure}

\begin{figure}
  \centering
  \includegraphics[scale=0.4]{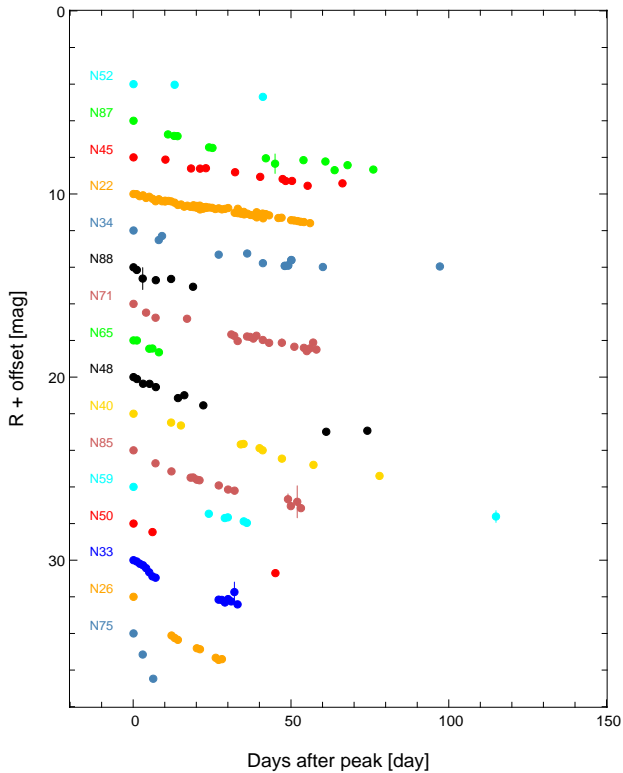}
  \caption{S Class novae with fixed $t_0$. The single offsets are $-$8.84 for N22, 15.17 for N26, 
12.44 for N33, $-$6.44 for N34, 4.97 for N40, $-$10.40 for N45, 2.53 for N48, 
10.92 for N50, $-$13.97 for N52, 8.05 for N59, $-$0.22 for N65, $-$1.84 for N71, 16.24 for N75, 6.70 for N85,
$-$11.06 for N87 and $-$5.14 for N88 respectively. 
Here we only show the decline part of the light curve. Full light curves can be found in the appendix.}
  \label{fig.s-class_fix_t0}
\end{figure}

\begin{table}
  \centering
    \caption{Power-law decline fitting for s-class nova}
    \begin{tabular}{lcc}
      \multicolumn{3}{c}{}\\
      \multicolumn{3}{c}{Free $t_0$}\\
      \hline
      Name & $t_0$(JD$-$2450000)       & $\alpha$         \\
      \hline
      N08 &  1337.9 $\pm$ 0.5 & $-$2.07 $\pm$ 0.02 \\
      N30 &  2257.3 $\pm$ 0.1 & $-$1.55 $\pm$ 0.01 \\
      N31 &  2277.9 $\pm$ 0.2 & $-$1.29 $\pm$ 0.02 \\
      N37 &  2647.4 $\pm$ 1.4 & $-$3.44 $\pm$ 0.28 \\
      N39 &  2776.6 $\pm$ 1.1 & $-$1.31 $\pm$ 0.04 \\
      N42 &  2831.0 $\pm$ 0.2 & $-$1.21 $\pm$ 0.03 \\
      N68 &  3627.7 $\pm$ 0.6 & $-$1.08 $\pm$ 0.05 \\
      N72 &  3845.5 $\pm$ 2.4 & $-$2.04 $\pm$ 0.16 \\
      N79 &  4088.9 $\pm$ 0.4 & $-$0.89 $\pm$ 0.03 \\
      N82 &  4155.1 $\pm$ 1.6 & $-$2.06 $\pm$ 0.23 \\
      N83 &  4288.2 $\pm$ 2.1 & $-$2.55 $\pm$ 0.51 \\
      N84 &  4289.8 $\pm$ 0.9 & $-$1.05 $\pm$ 0.08 \\
      N90 &  4437.3 $\pm$ 1.5 & $-$3.20 $\pm$ 0.44 \\
      \hline
      \multicolumn{3}{c}{}\\
      \multicolumn{3}{c}{Fixed $t_0$}\\
      \hline
      Name & $t_0$(JD$-$2450000)       & $\alpha$         \\
      \hline
      N22 & 1964.3 &  $-$4.35 $\pm$ 0.17 \\
      N26 & 2145.5 &  $-$1.92 $\pm$ 0.04 \\
      N33 & 2321.4 &  $-$1.22 $\pm$ 0.02 \\
      N34 & 2447.5 &  $-$1.70 $\pm$ 0.06 \\
      N40 & 2798.6 &  $-$1.67 $\pm$ 0.03 \\
      N45 & 2908.3 &  $-$0.97 $\pm$ 0.03 \\
      N48 & 2971.4 &  $-$0.92 $\pm$ 0.01 \\
      N50 & 2985.3 &  $-$1.37 $\pm$ 0.04 \\
      N52 & 3003.3 &  $-$1.43 $\pm$ 0.03 \\
      N59 & 3304.4 &  $-$1.63 $\pm$ 0.32 \\
      N65 & 3408.4 &  $-$1.44 $\pm$ 0.10 \\
      N71 & 3813.3 &  $-$2.92 $\pm$ 0.22 \\
      N75 & 3980.5 &  $-$2.28 $\pm$ 0.15 \\
      N85 & 4298.5 &  $-$1.16 $\pm$ 0.05 \\
      N87 & 4299.5 &  $-$1.22 $\pm$ 0.11 \\
      N88 & 4339.6 &  $-$1.48 $\pm$ 0.40 \\
      \hline
    \label{tab.free_t0}
    \end{tabular}
\end{table}

\subsection{C Class}

\begin{figure}[!h]
  \centering
  \includegraphics[scale=0.6]{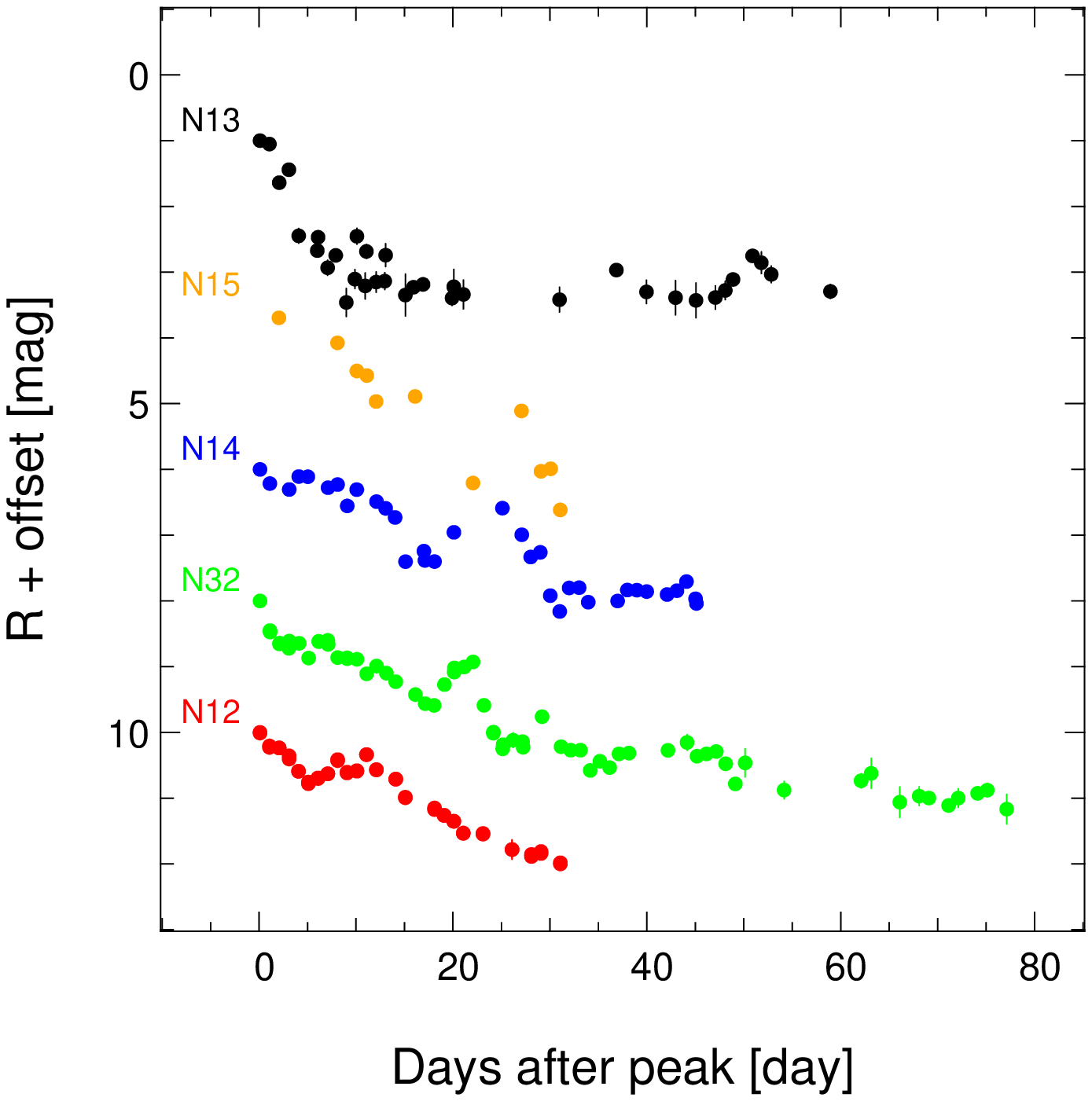}
  \includegraphics[scale=0.6]{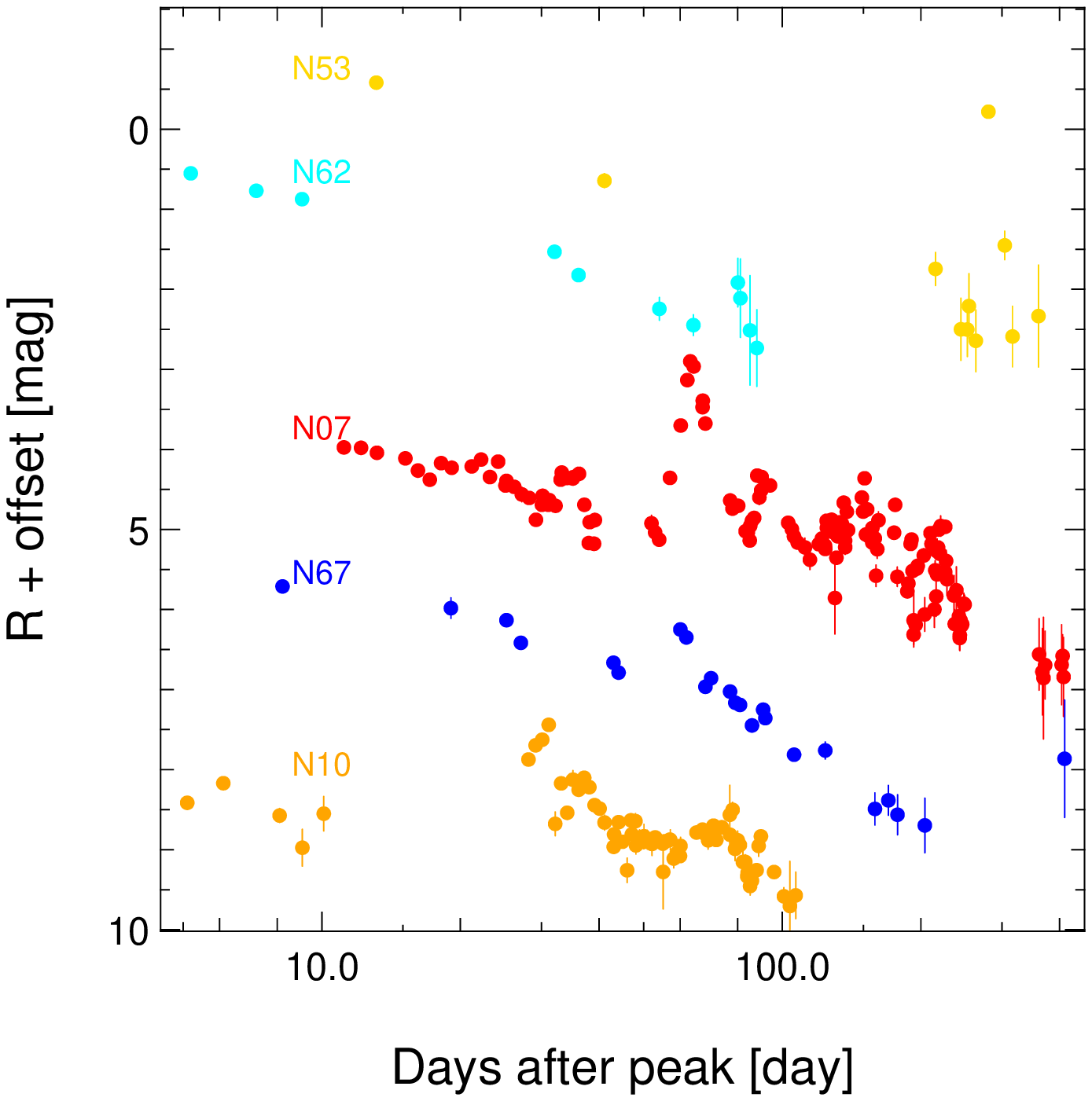}
  \caption{C Class novae. The offsets applied to the magnitudes are $-$15.07 for N07, $-$11.88 for N10, $-$8.26 for N12, 
$-$17.71 for N13, $-$11.58 for N14, $-$13.79 for N15, $-$9.16 for N32, $-$19.19 for N53, $-$19.15 for N62 
and $-$13.04 for N67 respectively. Here we only show the decline part of the light curve. 
Full light curves can be found in the appendix.}
  \label{fig.c-class}
\end{figure}

The light curves of C-class novae have cusp shape, which first follow a power-law decline, 
then rise steeply to a second maximum and finally have a sharp drop.
The characteristic C-class light curve has 
a secondary maximum emerging between 1 to 8 months after the primary peak \citep{2010AJ....140...34S}.
\cite{2009arXiv0904.2228K} found that C-class light curve can be well-fitted by an exponential 
component superimposed on the smooth decline. They further proposed that the cusps can originate 
from a secondary ejection and the break-out into the optically thick nova winds. 
\cite{2009ApJ...694L.103H} also connect the formation of the cusp shape to the input of the magnetic energy
from rotating white dwarf. In addition, the sharp drop before the light curve returns to the power-law decline is 
attributed to the sudden formation of dust as proposed by \cite{2008AJ....136.1815L}.
We have found in total 10 candidates showing cusp features in our WeCAPP catalog and show their light curves in 
Fig. \ref{fig.c-class}.

\subsection{O Class}

The O-class light curve follows the S-class light curve, but with the exception that 
at a given time interval the light curve shows quasi-sinusoidal self-similar oscillations during the course of decline.
It has been shown that the white dwarf of the O-class novae is both highly magnetic and massive 
\citep{2010AJ....140...34S}. However, these can not be the only effect leading to oscillation because 
the nova V1500 Cyg in \cite{2010AJ....140...34S} which fulfills these requirements but does not show 
oscillation. There have been many proposals for the mechanism of oscillation, but none of them have been 
compared to and verified by observation \citep[see Sect. 4 in ][]{2010AJ....140...34S}. 
The oscillation starts generally around 3 mag below the peak, which indicates
that we might have missed the peak in our nova candidates N11 and N28, where the light curves are shown in 
Fig. \ref{fig.o-class} and in the appendix. 
In Fig. \ref{fig.o-class} we show the two O-class candidates discovered during our observation campaign.

\begin{figure}[!h]
  \centering
  \includegraphics[scale=0.41]{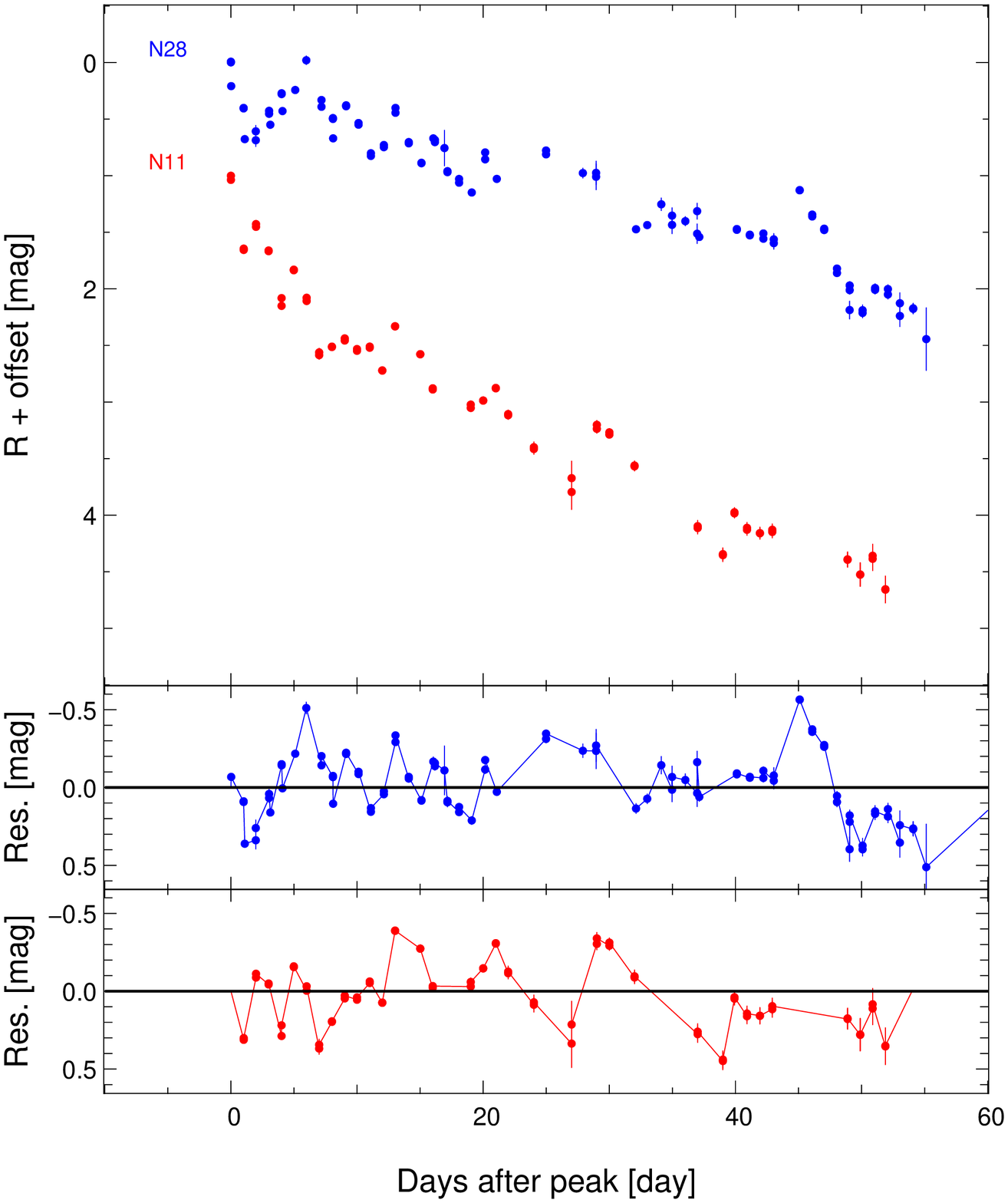}
  \caption{O Class novae. The offset is $-$15.77 for N11 and $-$17.61 for N28 respectively. 
Here we only show the decline part of the light curve. Full light curves can be found in the appendix.}
  \label{fig.o-class}
\end{figure}

\subsection{J Class}

\begin{figure}[!h]
  \centering
  \includegraphics[scale=0.41]{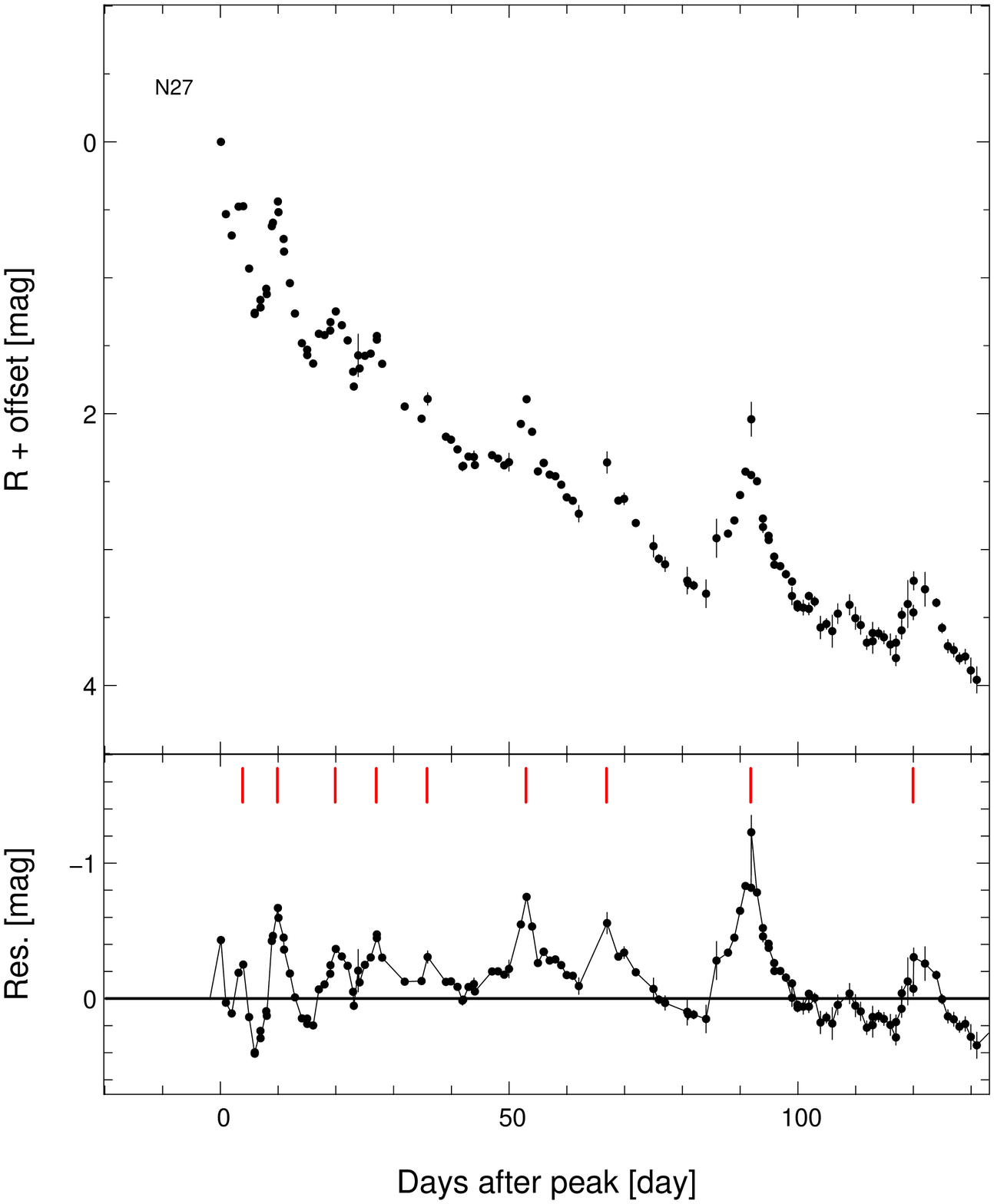}
  \caption{J Class nova. The single offsets are $-$16.99 for N27. 
Here we only show the decline part of the light curve. Full light curves can be found in the appendix.}
  \label{fig.j-class}
\end{figure}

The characteristics of J-class novae are the jitters on top of the smooth decline. 
These jitters are symmetric and sharp-topped flares superposed on the base of S-class light curve, 
which is the major difference from the O-class novae, while the latter bears oscillations 
up and down the smooth decline. The jitter usually has variations with amplitude larger than half of 
a magnitude. Jitters do not occur in the late tail of the light curve and most of them occur within
3 mag below the peak. \cite{2010AJ....140...34S} further propose for two subclasses according to the 
emergence of the jitters: one subclass has jitters only near the peak, while the other has jitters 
spread all over the light curve roughly until the nova is 3 mag dimmer than the peak. 
Among our candidates we found one evident J-class light curve, which belongs to the second subclass 
of \cite{2010AJ....140...34S} and is shown in Fig. \ref{fig.j-class}.

It has been reported that there is a gradual increase of the time intervals between two successive jitters
\citep{1992A&A...257..599B, 2005A&A...429..599C, 2009ApJ...701L.119P, 2010arXiv1010.5611T}, while 
\cite{2010AJ....140...34S}, using the same data set as \cite{2009ApJ...701L.119P}, found no distinctive trend. We thus 
tried to search for such trend in our nova candidate N27 and performed a fitting with the following equation:
\begin{equation}
\log (t_J - t_{J-1}) = a\log (t_J-t_{max}) + b,
\end{equation} 
where $t_J$ is the time of the $J$-th jitter.
\begin{figure}
  \centering
  \includegraphics[scale=0.6]{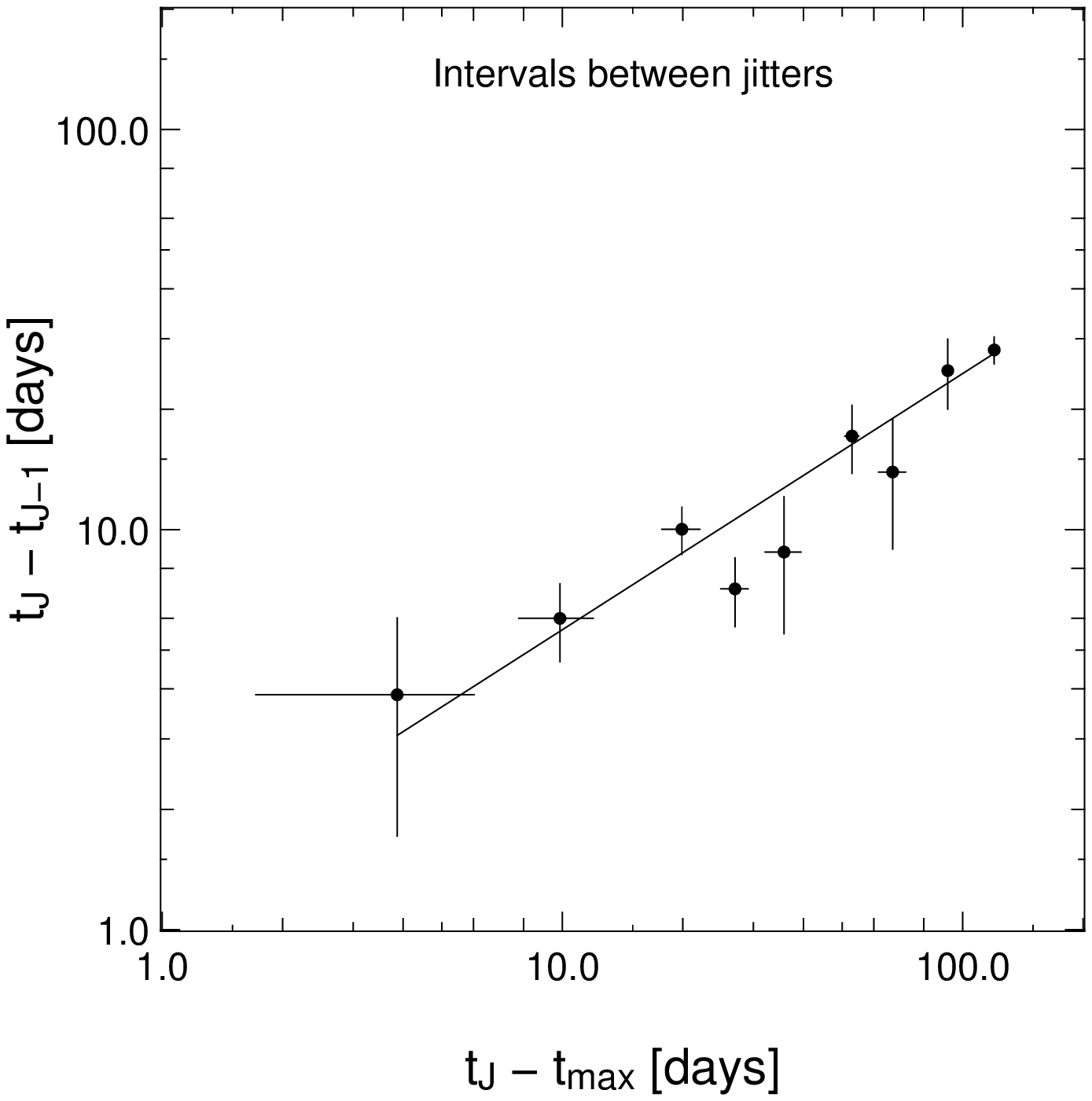}
  \caption{J-class peak intervals for nova N27, using $R$-band data.}
  \label{fig.j-peak}
\end{figure}

The jitters used in the fitting are indicated by the vertical marks in Fig. \ref{fig.j-class}. The time intervals between the 
successive jitters are shown in \ref{fig.j-peak}. Our best-fitted
value is $a$ = 0.64$\pm$0.09 and $b$ = 0.11$\pm$0.16. The slope is smaller 
than the values of DK Lac ($a$ = 0.88) and V4745 Sgr ($a$ = 0.79) derived by \cite{2009ApJ...701L.119P} and a = 0.79 for the 
6 novae presented by \cite{2010arXiv1010.5611T}. With only one J-class nova candidate in our catalog, we can not 
tell if this is a difference between the nova in M31 and Galactic novae, or it is simply a variation among individual 
novae. 

\subsection{Other classes}
Besides the above-mentioned classes, there remains three more classes in the taxonomy of \cite{2010AJ....140...34S}:
\begin{itemize} 
\item Flat topped (F) class which has an extended interval at the peak with near constant brightness. 
\item Dust dip (D) class where the decline is interrupted by another very steep decline and followed by the 
recovery to just below the original decline. 
\item Plateau (P) class that the smooth decline is interrupted by a long-lasting and nearly flat interval, 
succeeded by the return to the original decline. 
\end{itemize}
Among our candidates, however, we do not find evident light curves belonging to these classes. This could be partially 
attributed to the set-up of our observation campaign. For example, the dust dip for the extreme shallow dips 
in \cite{2010AJ....140...34S} occur more than 1 month after the peak, with the dip to be about 6 mag dimmer than 
the light curve maximum. Such magnitude variation can hardly be observed in M31, because it is too faint to be 
discerned. This implies that we might misclassified the D-class novae into other classes. The non-detection of the P-class novae can be explained by the filter system we used. 
\cite{2006ApJS..167...59H} pointed out that the true plateau from the continuum radiation is best observed in 
the \textit{y}-band filter. Since we are using the \textit{R} and \textit{I}-filter, it is possible that the 
plateau phase does not exist in the $R$ and $I$-bands due to the influence of the emission lines 
during the course of decline. 

To summarize, we have classified 42 nova candidates and find 69\% to be S-class, 24\% to be C-class, 5\% to be O-class and 2\% to be J-class, while 
\cite{2010AJ....140...34S} find 38\% to be S-class, 1\% to be C-class, 4\% to be O-class and 16\% to be J-class in their sample.

\section{Recurrent Novae}
\footnotesize
\begin{table*}[t]
  \centering
  \caption{Recurrent Nova candidates}
  \begin{minipage}{165mm}
    \begin{tabular}{cccccccc}
      \hline
      WeCAPP ID           & RA          & DEC         & NAME          & RA          & DEC         & Separation & $\Delta_{M31}$ \\
      \hline
      N02 (M31N-1997-10f) & 00:42:52.35 & +41:16:13.2 & M31N-2008-08b & 00:42:52.38 & +41:16:12.9 & 0$\farcs$54      & 1$\farcm$50\\
      N19 (M31N-2001-07b) & 00:42:57.75 & +41:08:12.3 & M31N-1963-09c & 00:42:57.73 & +41:08:12.4 & 0$\farcs$32      & 8$\farcm$29\\
      N19 (M31N-2001-07b) & 00:42:57.75 & +41:08:12.3 & M31N-1968-09a & 00:42:57.71 & +41:08:11.9 & 0$\farcs$72      & 8$\farcm$29\\
      N19 (M31N-2001-07b) & 00:42:57.75 & +41:08:12.3 & M31N-2010-10e & 00:42:57.76 & +41:08:12.3 & 0$\farcs$15      & 8$\farcm$29\\
      N29 (M31N-2001-12b) & 00:42:39.59 & +41:09:02.9 & M31N-1997-11k & 00:42:39.59 & +41:09:02.9 & 0$\farcs$00      & 7$\farcm$12\\
      N29 (M31N-2001-12b) & 00:42:39.59 & +41:09:02.9 & M31N-2009-11b & 00:42:39.61 & +41:09:03.2 & 0$\farcs$42      & 7$\farcm$12\\
      N83 (M31N-2007-07a) & 00:43:04.05 & +41:17:08.3 & M31N-1990-10a & 00:43:04.05 & +41:17:07.5 & 0$\farcs$80      & 3$\farcm$82\\
      \hline
    \end{tabular}
    \tablefoot{We give the WeCAPP name, the positions (also see 
        Fig. \ref{fig.all}), the corresponding novae 
        fulfilling the 1$\farcs$0 criterion, the separation and the distance from the center of M31 ($\Delta_{M31}$).}
    \label{tab.recurrent-nova}
  \end{minipage}
\end{table*}
\normalsize

\begin{figure}[!h]
  \centering
  \includegraphics[scale=0.45]{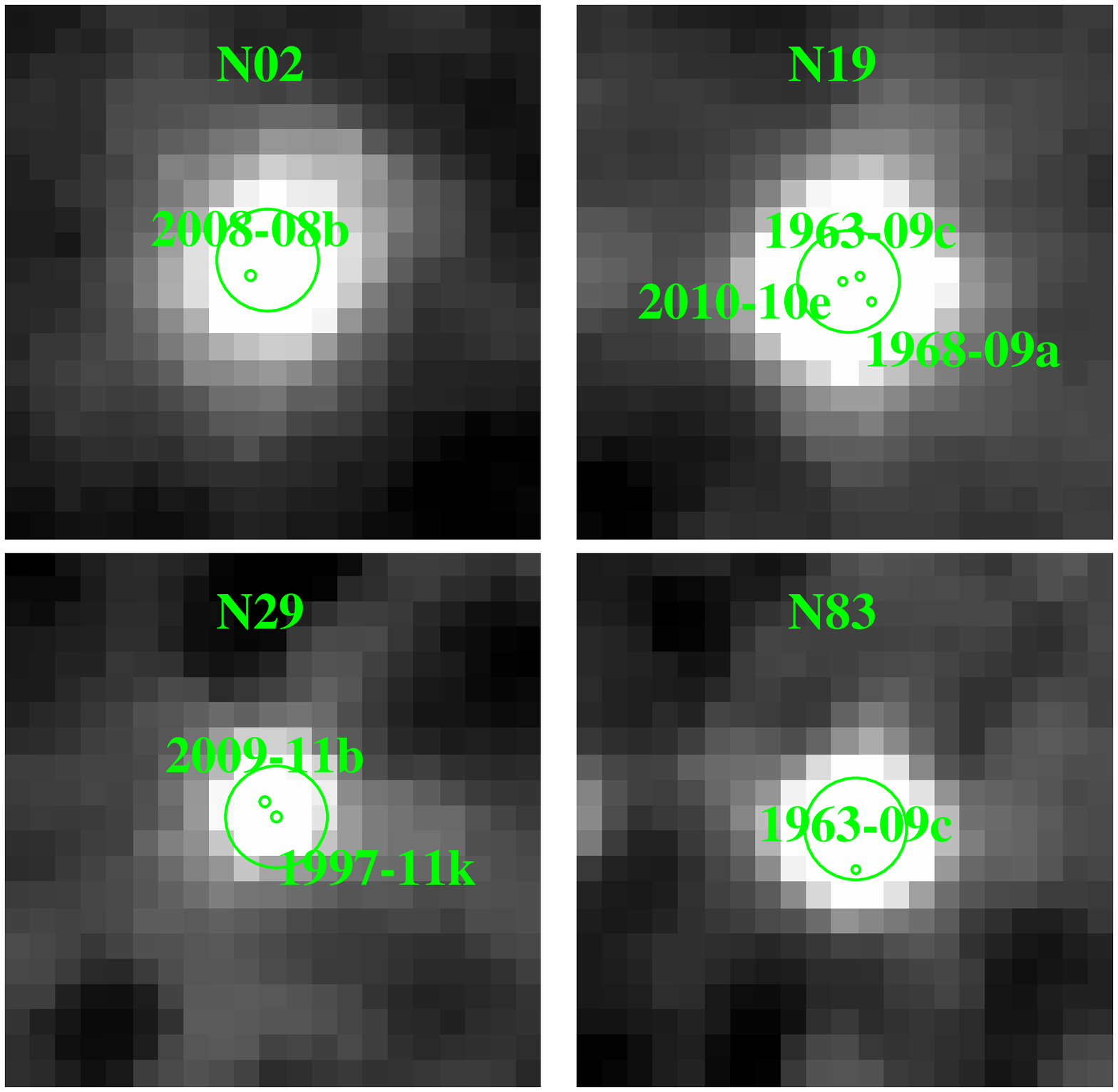}
  \caption{Position of the recurrent. The larger circle in the center indicates the 1$\farcs$0 radius for our selection criteria. The position 
    of potential recurrent nova candidates are marked by the smaller circles.}
  \label{fig.RNe}
\end{figure}

Recurrent novae are potential supernovae progenitors \citep{2010ApJS..187..275S}. We compare the position of 
our nova candidates 
with the catalog by \cite{2007A+A...465..375P, 2010AN....331..187P}.
We have found 4 recurrent novae candidates by selecting novae in the literature which are located within
1 arcsec to our nova candidates (see Table \ref{tab.recurrent-nova} and Fig. \ref{fig.RNe}). 
Among the potential recurrent nova candidates N29 has 3 outbursts in 12 years, which would be an unprecedented short period.
As pointed out by \cite{2009ATel.2286....1H}, the outburst appears earlier in UV and H$\alpha$ than in the $R$-band which does not 
fit very well to the nova scheme. They thus suggest an alternative scheme, that this event could be a dwarf nova in the Milky Way.
N19 has 4 outbursts detected so far. Because the short time separation between the first two outbursts, \cite{1989SvAL...15..382S} have doubted 
its nova nature and suggested it to be a U Gem type foreground Galactic dwarf nova. However, the spectroscopic observations of the 4th outburst 
in 2010 \citep{2010ATel.3006....1S} have confirmed it as a He/N spectroscopic class nova located in M31. In addition, \cite{2010ATel.3038....1P}
have also reported the SSS turn-on $\sim$ 15 days after the first optical detection in 2010.

To test how likely an uncorrelated nova is falling into the 1 arcsec area, we perform a test by using the upper-right quarter of our 
pointing F1, which has the highest M31 light contribution from the bulge and contains 42 novae. The ratio of the area occupied by 
the 1$\farcs$0 circle of these 42 novae to the total area of this quarter (300$\times$300 arcsec$^2$), implies the chance of an uncorrelated 
nova to coincide with an existing nova is low (1.5 : 1000).
As most of the recurrent novae are not found in this quadrant (see Fig. \ref{fig.all} and Table \ref{tab.recurrent-nova}),
the chance of coincidence is even smaller for the majority of the recurrent nova candidates.

Note that we use stricter selection criteria to search for recurrent novae, thus we have less candidates than 
presented by \cite{2007A+A...465..375P, 2010AN....331..187P}. 

\cite{2006ApJS..167...59H} suggested that recurrent novae all bear the plateau light curve. However, 
in our light curve we did not detect evident plateaus. The main reason is we do not have comprehensive coverage 
of the light curves. 
Despite of the lack of highly sampled observation, it would be hard to find such plateaus because the 
light curves 
in \textit{R} or \textit{I} are contaminated by the bright emission lines. \cite{2008ASPC..401..206H} thus 
advocate observations in Str\"omgren $y$-band (centered at 547 nm) since it is designed to cut the strong emission lines in the 
wide $V$ bandpass filter and can follow the continuum flux more accurately. However, the Str\"omgren $y$-filter is narrow 
and requires longer exposure time, so we use the $I$-filter instead of the Str\"omgren $y$-filter for the confirmation of 
microlensing event from achromaticity when the WeCAPP was initiated. 

\begin{figure}[!h]
  \centering
  \includegraphics[scale=0.56]{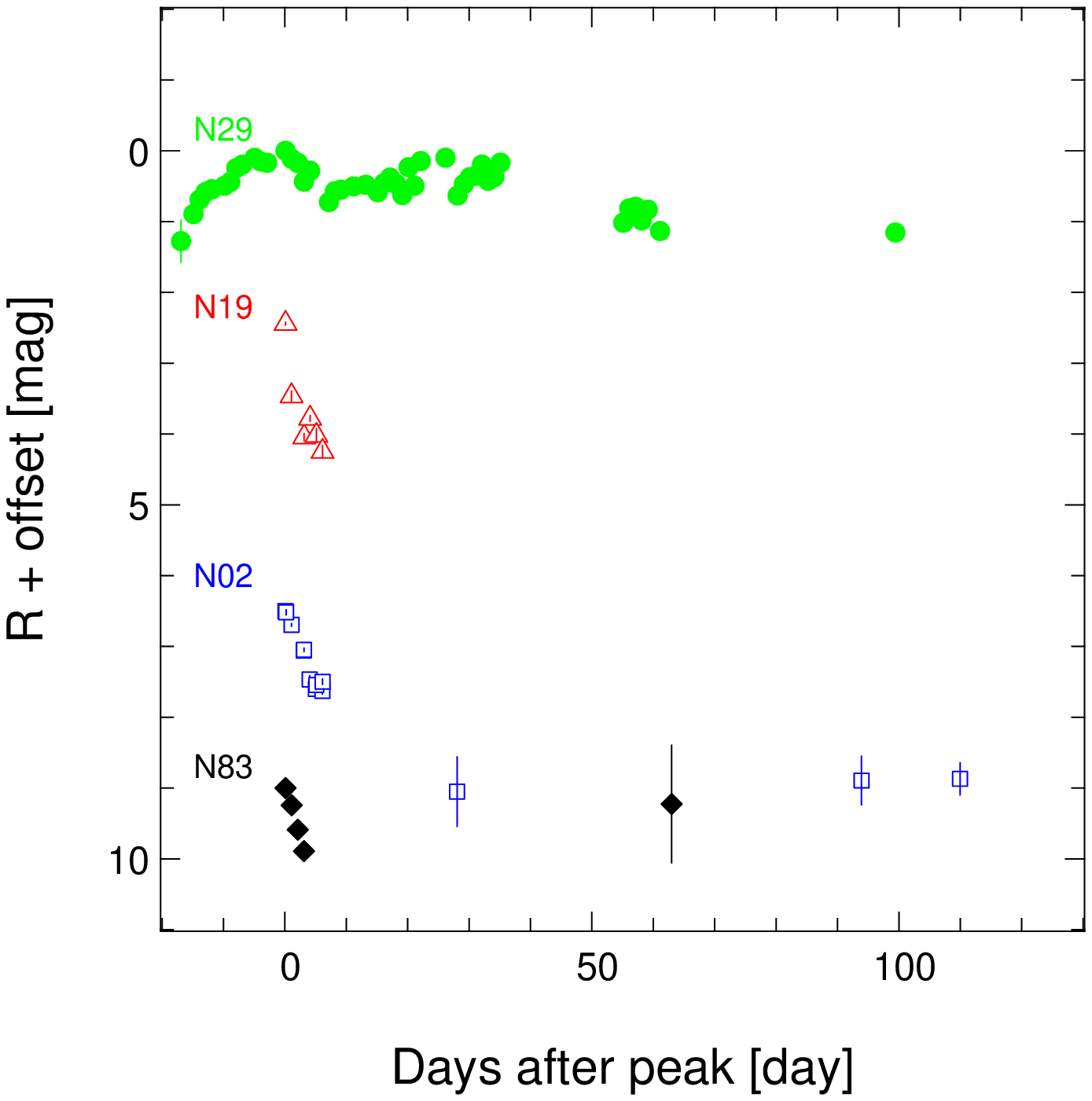}
  \caption{Recurrent nova candidates light curves. The single offsets are $-$11.76 for N02, $-$16.76 for N19,  
$-$18.86 for N29 and $-$9.86 for N83 respectively. Here we only show the decline part of the light curve. 
Full light curves can be found in the appendix.}
  \label{fig.rn_lc}
\end{figure}

\section{Rate of decline and the speed class}
\begin{table*}[!th]
\centering
\caption{Speed class of nova according to \cite{1989clno.conf....1W}}
\begin{tabular}{llrccc}
\hline
Speed class     & $t_2$       & d$V$/d$t$   & \multicolumn{2}{c}{M31 sample}  &  MW sample         \\
                & [day]       & [mag/day]   & This work  & \cite{2004MNRAS.353..571D} & \cite{2010AJ....140...34S}  \\
\hline 
Very fast       & $\le$ 10    & $\ge$ 0.2   &  8      & 1           & 35                          \\
Fast            & 11-25       & 0.18-0.08   &  18     & 3           & 27                          \\
Moderately Fast & 26-80       & 0.07-0.025  &  46     & 11          & 23                          \\
Slow            & 81-150      & 0.024-0.013 &  11     & 2           & 7                           \\
Very slow       & 151-250     & 0.013-0.008 &  7      & 3           & 3                           \\
                &             & $\le$0.08   &  1      & 0           & 2                           \\
\hline
\label{tab.speed}
\end{tabular}
\end{table*}

\begin{figure}
  \centering
  \includegraphics[scale=0.5]{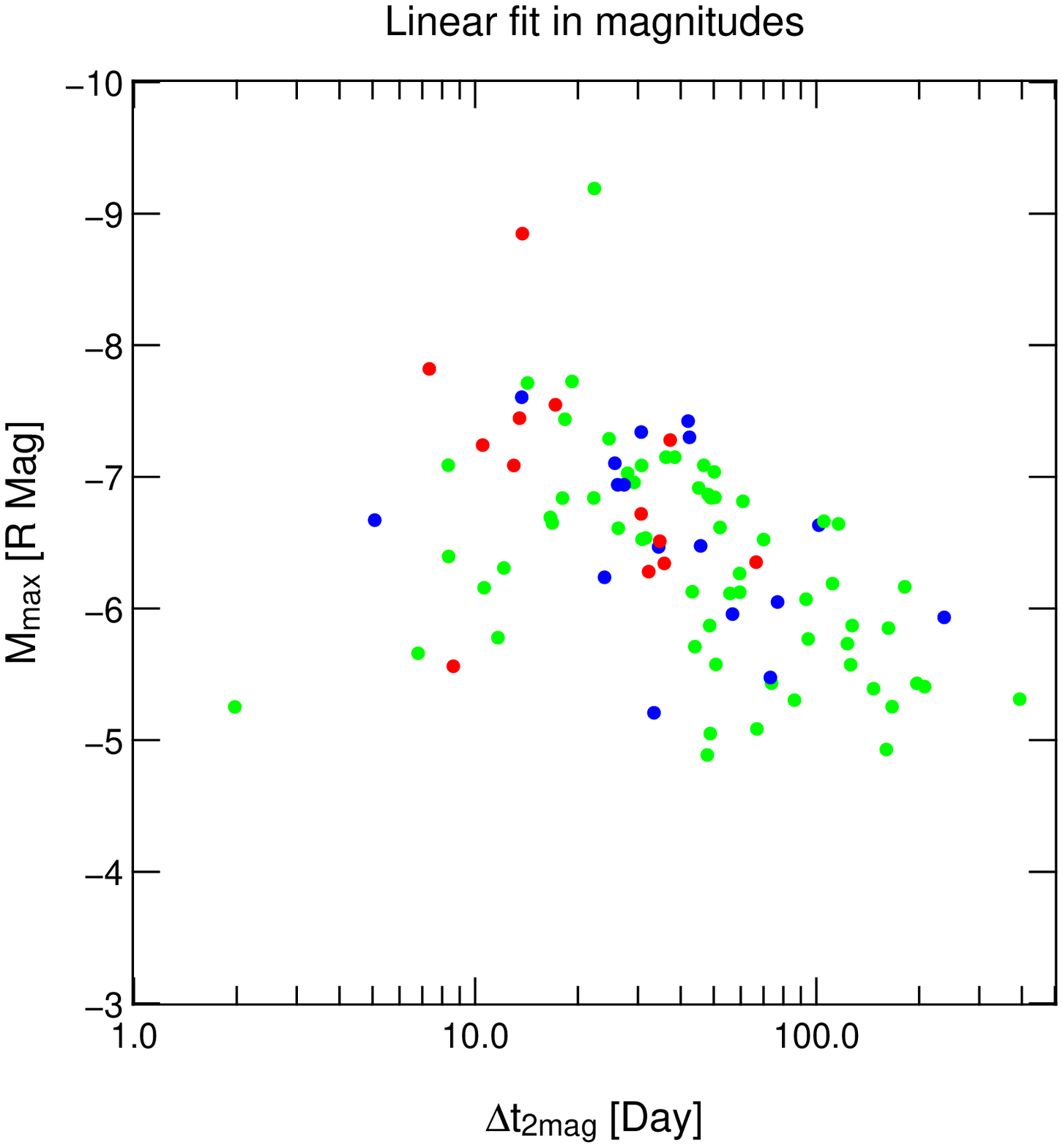}
  \caption{Distribution of the observed apparent maximum brightness in $R$-band and the 
      fitted $t_2$ for all nova candidates. The red and blue points are referred to equation (\ref{eq.free_t0}) 
      with $t_0$ as free parameter and equation (\ref{eq.fixed_t0}) with
      $t_0$ as fixed parameter, respectively. The green points are novae belong to other classes.
      The $t_2$ value is derived from the dm/dt and 
      the observed apparent maximum in the light curves. See the main text for detailed description.}
  \label{fig.mmrd_all}
\end{figure}

\begin{figure}
  \centering
  \includegraphics[scale=0.5]{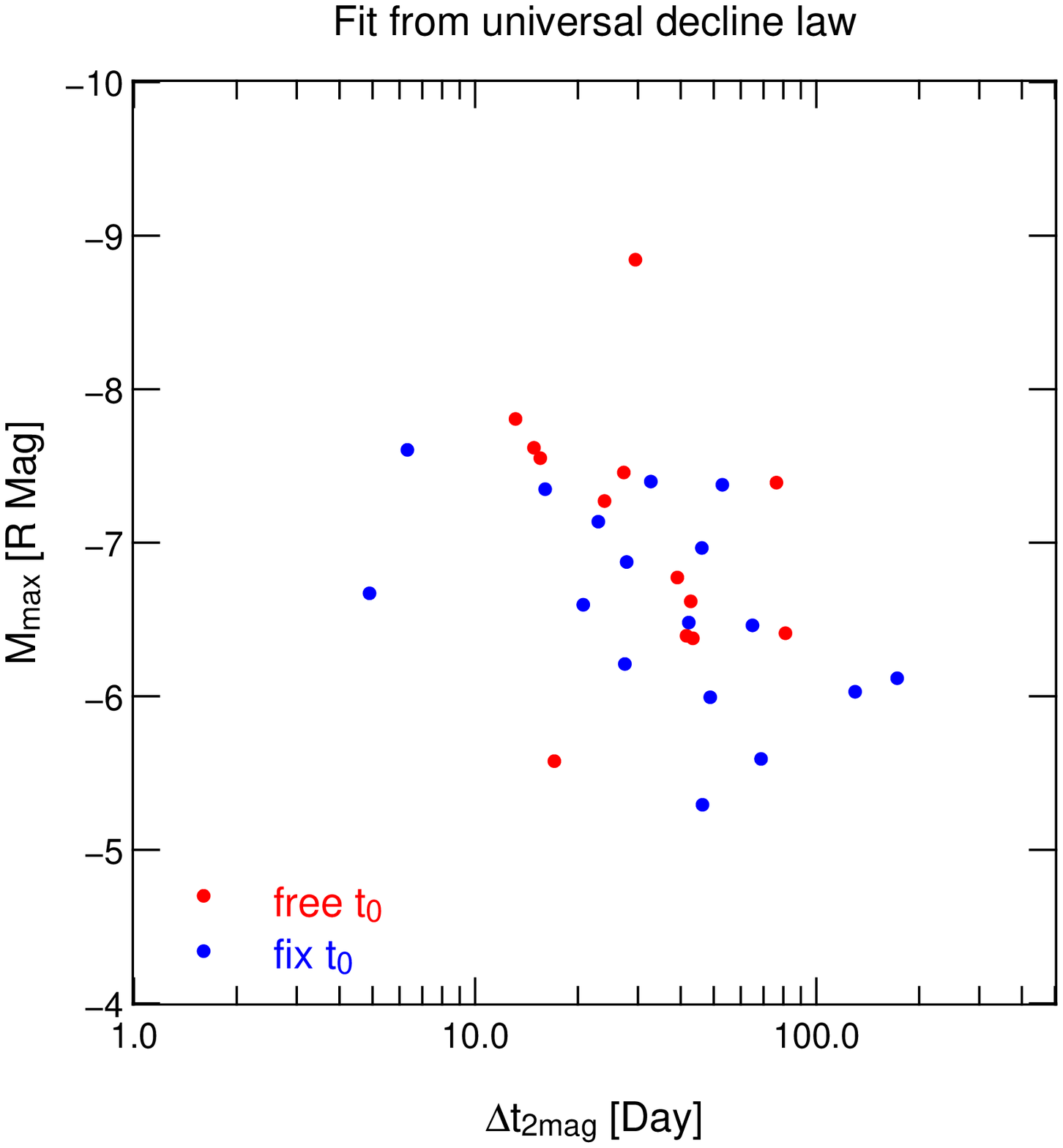}
  \caption{Distribution of the observed apparent maximum and the fitted $t_2$ for S-class novae. The $t_2$ value is derived from 
    the universal decline law \citep{2006ApJS..167...59H} and the observed apparent maximum in the light curves. See the main text 
    for detailed description. The red and blue points are referred to equation (\ref{eq.free_t0}) with $t_0$ as free parameter and equation (\ref{eq.fixed_t0}) with
    $t_0$ as fixed parameter, respectively.}
  \label{fig.mmrd_s-class}
\end{figure}

In this section we present the rate of decline for our nova sample. Due to the observing gaps before 
some of the apparent maxima, it is hard to recover the true maximum fluxes during the nova eruption. Nevertheless,
the apparent maxima can serve as a lower limit of the true maxima. One can derive upper limits for the 
$t_2$ values (the time required for the nova to faint by two magnitudes) if the apparent maxima are taken as the true maxima.
We have retrieved the $t_2$ values for our sample as follows: we perform a linear fitting on the decline part of 
each light curve and determine the rate of decline, dm/dt (in units of magnitude/day). We then use dm/dt to 
derive the $t_2$ value relative to the observed maximum magnitude for all the novae. The result is shown in Fig. \ref{fig.mmrd_all}.
Besides the linear fitting, we also applied the universal decline law proposed by ~\cite{2006ApJS..167...59H} to 
retrieve the $t_2$ value from the observed magnitude at the apparent maximum. This procedure can only be done for 
30 S-class novae, because the fitting routine fails to find a solution for other classes. The result is shown in 
Fig. \ref{fig.mmrd_s-class}. The reader should keep in mind, that Fig. \ref{fig.mmrd_all} or \ref{fig.mmrd_s-class} 
does not give the exact MMRD relation, but serves as an upper limit in the $t_2$ and lower limit in the magnitude.

\begin{figure}
  \centering
  \includegraphics[scale=0.5]{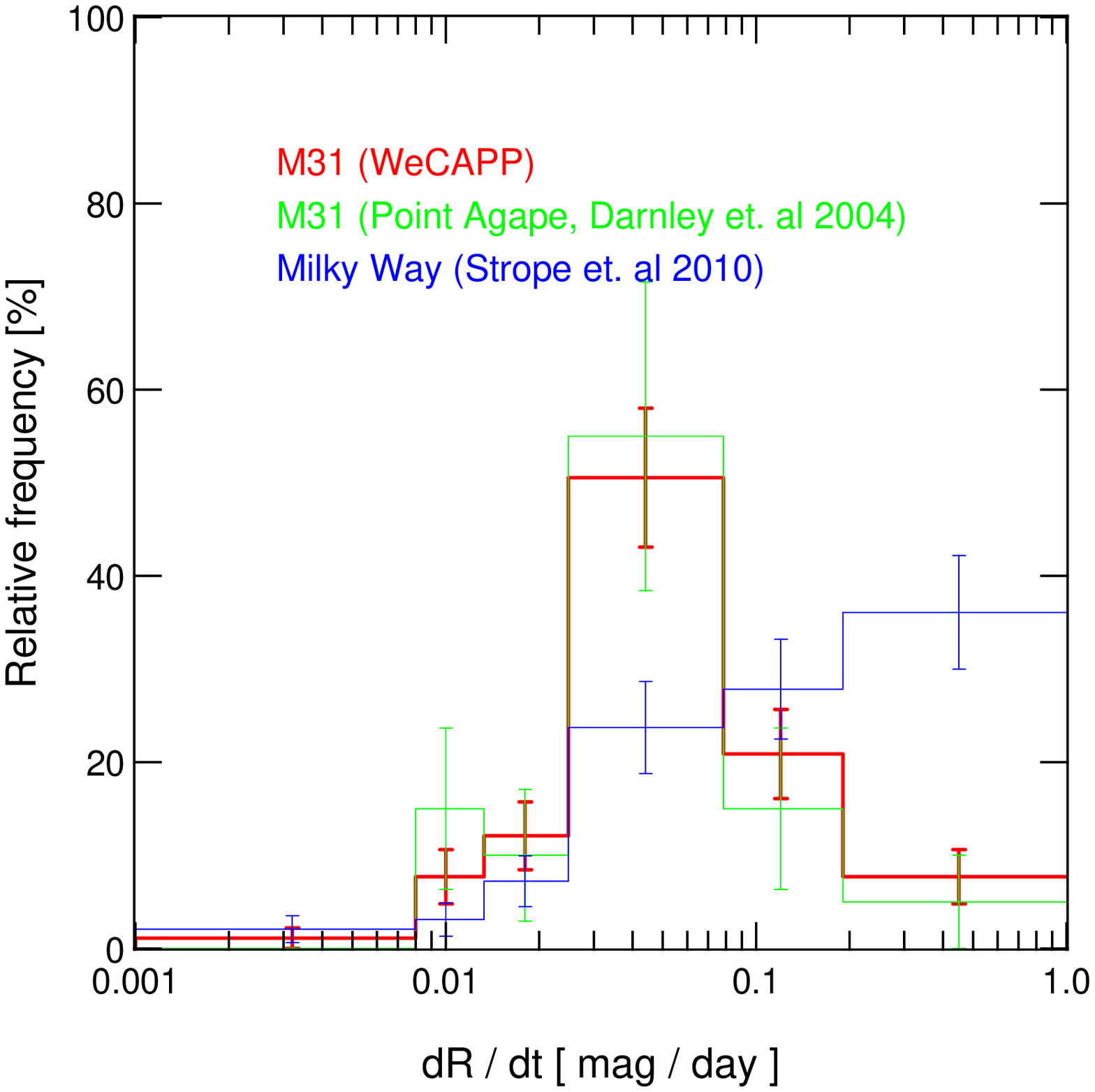}
  \caption{The distribution of the speed class of novae in M31 (see Table \ref{tab.speed}, 
        for definition see \cite{1989clno.conf....1W}). The red line is derived from our sample and the green line is from 
      the M31 novae presented by \cite{2004MNRAS.353..571D}. The blue line represents the Milky Way novae by \cite{2010AJ....140...34S}.}
\label{fig.n_fq}
\end{figure}

As pointed out by \cite{1989clno.conf....1W}, the value of dm/dt is also an indicator of the speed class. We thus use the dm/dt values 
of our sample to derive the frequency of different speed classes of nova in M31. A comparison of the dm/dt distribution
in our sample to the novae presented by \cite{2004MNRAS.353..571D} is shown in Fig. \ref{fig.n_fq}, the two M31 samples agree rather well
within the statistical errors (shown are only the $\sqrt{n}$ number count errors). The Milky Way data set of \cite{2010AJ....140...34S} 
differs significantly by presenting a much higher fraction of very fast novae than the M31 data. For a fair comparison, one would 
have to correct the \cite{2010AJ....140...34S} sample for its severe observational selection effects as being observed from inside the 
(dusty) Milky Way disk. A more detailed comparison is beyond the scope of this paper as it requires detailed modeling of the distribution 
of stellar and dust population of both galaxies.

\section{Conclusion and outlook}
We have presented the position, outburst time and the maximum brightness of the 91 nova candidates
discovered during the time span of the WeCAPP project. Light-curve classifications under the taxonomic 
scheme of \cite{2010AJ....140...34S} have been shown and the full \textit{R} and \textit{I}-band 
light curves of each individual nova during the outburst are also presented in the Appendix. 

In this work we successfully applied a scheme developed for a Milky Way nova
sample which, because of observational selection effects, is certainly dominated by the galactic disk, 
to a nova sample of a different host galaxy which is mostly observed towards the bulge of this host. The
differences in member ratios for the subclasses as defined by \cite{2010AJ....140...34S} between the 
Milky Way and our M31 WeCAPP sample probably reflect to some extent different observational selection 
effects, but bear the potential for further conclusions on the differences between stellar populations 
in M31 and the Milky Way, once the selection effects are proper accounted for.

We provide the full light curve data of the novae on request, as well as
the postage stamps of the reduced, stacked, or difference-imaging frames.

Part of this catalogue has been used to find the X-ray counter-part by \cite{2005A+A...442..879P, 2007A+A...465..375P} 
and showed that super soft X-ray sources (SSS) in M31 are mostly constituted by the novae during eruption. The turn on 
and turn off of the SSS phase provide us the information of the ejected and accreted mass onto the surface
of the white dwarf.   

Besides the X-ray monitoring campaign, there is also a survey of M31 novae in infrared using 
\textit{Spitzer Space Telescope} \citep{2011ApJ...727...50S}, which indicates a correlation 
between the dust formation timescales and the nova
speed class. Such studies would not be possible without the speed class determined by the optical observations. 
Ground-based optical surveys, such as PTF \citep{2009PASP..121.1395L,2009PASP..121.1334R}, 
PanSTARRS \citep{2002SPIE.4836..154K} and LSST \citep{2002SPIE.4836...10T}, will continue to play 
an important role in the regime of multi-wavelength novae observation and help us to gain insight of the 
underlying physical mechanism of novae.  

\begin{acknowledgements} 
We are grateful for the comments from the anonymous referee. We thank Sara B\"uhler and Silona Wilke for their contributions in observation. 
This work was supported by the DFG cluster of excellence `Origin and Structure of the Universe' (www.universe-cluster.de).
\end{acknowledgements}

\bibliographystyle{aa}
\bibliography{literature}

\end{document}